\documentclass[aps,12pt,final,oneside,onecolumn,nobibnotes,nofootinbib,% 
superscriptaddress,centertags,floatfix,secnumarabic,notitlepage]{revtex4}%

\usepackage{bm}
\usepackage{graphics}
\usepackage{rotating}
\usepackage{epsfig}
\usepackage{amsmath}
\usepackage{amsfonts}
\usepackage{varioref}
\usepackage[hidelinks]{hyperref}

\begin{document}

\title{Ground state hyperfine structure of light \\ muon - electron ions}

\author{R.~N.~Faustov\footnote{E-mail:~faustov@ccas.ru}}
\affiliation{Institute of Cybernetics and Informatics in Education, FRC CSC RAS, Moscow, Russia}

\author{V.~I.~Korobov\footnote{E-mail:~korobov@theor.jinr.ru}}
\affiliation{Bogoliubov Laboratory of Theoretical Physics JINR, Dubna, Russia}

\author{A.~P.~Martynenko\footnote{E-mail:~a.p.martynenko@samsu.ru}}
\affiliation{Samara National Research University, Samara, Russia}

\author{F.~A.~Martynenko\footnote{E-mail:~f.a.martynenko@gmail.com}}
\affiliation{Samara National Research University, Samara, Russia}

%\date{\today}

\begin{abstract}
The ground state hyperfine splitting of light muon - electron ions of lithium, beryllium, boron and helium is 
calculated on the basis of analytical perturbation theory in terms of small parameters of the fine 
structure constant and electron - muon mass ratio. The corrections of vacuum polarization, nuclear 
structure and recoil effects and electron vertex corrections are taken into account. The dependence 
of the corrections on the nucleus charge Z is studied. The obtained total values 
of hyperfine splitting intervals can be used for comparison with future experimental data.
\end{abstract}

\maketitle

\section{Introduction}\label{Intro}

The precision investigation of fine and hyperfine structure of the simplest atoms, gyromagnetic 
factors of bound leptons, as well as the particle bound states production and decay processes makes 
it possible to test quantum electrodynamics (QED) and the relativistic theory of bound states. 
The precision muonic physics has become especially actual since 2010, when the first experimental 
results of low - lying energy levels measurement in muonic hydrogen were obtained by the CREMA collaboration 
(Charge Radius Experiments with Muonic Atoms). A decade of active work of this collaboration has brought 
interesting and unexpected results, related primarily to determining more accurate values of the charge 
radii of light nuclei (proton, deuteron, helion, alpha particle) \cite{crema1,crema2,crema3,crema4,crema5}:
\begin{equation}
\label{ep1}
\begin{cases}
r_p=0.84087(26)_{exp}(29)_{theor} ~fm,\\
r_d=2.12718 (13)_{exp}(89)_{theor} ~fm,\\
r_\alpha=1.67824(13)_{exp}(82)_{theor} ~fm.
\end{cases}
\end{equation}

As a result of the first experiments of the CREMA collaboration in 2010, value of proton charge radius 
$r_p=0.84184 (67)$ fm, was obtained. It was 10 times more accurate, than all previous values from 
experiments based on electronic systems. Moreover it was essentially smaller, than the CODATA recommended 
value $r_p=0.8768 (69)$ fm \cite{codata}. The difference between these values was named "proton radius puzzle". 
The measurements of energy levels with muonic hydrogen have shown that there is significant discrepancy in values 
of proton and deuteron charge radii, emerging from experiments with electronic and muonic atoms. 
New problems of studying the fine and hyperfine structure of the energy spectrum are related to muonic ions 
of lithium, beryllium, etc. \cite{schmidt}.

The CREMA experiments caused a series of new experimental studies of that problem. During 2017-2019 
different experimental results were obtained, both with electron and muon systems, which made 
it possible to refine the value of the proton charge radius.
The transition frequency (2S-4P) in electronic hydrogen was measured in \cite{beyer}: 
$\Delta \nu_{2S-4P}=616520931626.8(2.3)$ kHz, and extracted value of proton charge radius $r_p=0.8335(95)$ fm 
found to be in agreement with the CREMA result. 
Another experimental investigation of the "proton radius puzzle", PRad (E12-11-1062), was planned in 2011 
and successfully carried out in 2016 in the Thomas Jefferson National Accelerator Facility. PRad experiment 
was based on studying of electron beams with energy 1.1 and 2.2 GeV. In that experiment the cross section 
of the e - p elastic scattering at unprecedentedly low values of the square of the transferred momentum 
was measured up to a percentage. The obtained value of proton charge radius was 
$r_p = 0.831\pm 0.007(stat)\pm 0.012(syst)$ fm \cite{xiong}. It is less, than average of $r_p$ 
from previous elastic e - p scattering experiments, but agrees with the spectroscopic results 
for the muonic hydrogen atom within experimental uncertainties. 
A new measurement of the electronic hydrogen Lamb shift (n=2) was made in \cite{bezginov}. 
The result is: $\Delta E^{Ls}=909.8717(32)$ MHz. Value of proton charge radius from this experiment 
$r_p=0.833(10)$ agrees with spectroscopic data for muonic atoms. 
It should be noted that in another experiment \cite{fleurbaey} a new measurement of the two - photon 
transition frequency (1S - 3S) was measured with relative uncertainty $9 \cdot 10^{-13}$: 
$\Delta \nu_{1S-3S} = 2922743278671.0(4.9)$ kHz. The value of proton charge radius from this 
experiment $r_p = 0.877(13)$ fm is in good agreement with the recommended CODATA value. 
To solve the proton charge radius problem, the PSI MUSE collaboration is planning an experiment 
to simultaneously measure the cross sections for electron and muon scattering by protons \cite{gilman}. 
This experiment will make it possible to determine the charge radii of the proton independently 
in the two reactions and test the lepton universality with an accuracy of an order of magnitude better 
than previous scattering experiments.
At the J-PARC (Muon Science Facility (MUSE)) research center, the MuSEUM collaboration (Japan) plans 
an order of magnitude more accurate measurement of the hyperfine structure of the muonium ground state 
\cite{jparcmuse}. Another experiment of the MU-MASS at PSI (Switzerland) aims to measure the (1S-2S) 
transition frequency in muonium with an accuracy of 10 kHz (4 ppt) \cite{crivelli}.
New plans for precision microwave spectroscopy of the J-PARC MUSE collaboration \cite{strasser} involve 
measuring the hyperfine structure (HFS) of the ground state of muonic helium with an accuracy two orders 
of magnitude better than previous experiments in the 1980s. 
The FAMU (Fisica degli Atomi Muonici) collaboration plans to measure the hyperfine structure of the ground 
state of muonic hydrogen with an accuracy of several ppm \cite{pizz}, and the CREMA \cite{crema2021} 
collaboration with an accuracy of 1 ppm using pulsed laser spectroscopy methods.
Fundamental experiment related to muon physics is the Fermilab (USA) experiment on measuring the 
muon anomalous magnetic moment \cite{abi}, which recently confirmed that there is a difference of 4.2 
standard deviations between the experimental and theoretical values of the muon anomalous magnetic moment, 
which can be an indication of the New Physics beyond the Standard Model. Another project with the same goal, 
but delivered according to a completely different methodology, is planned to be carried out at J-PARC (Japan) 
\cite{abe}. All these experiments, already carried out and planned for the near future, convincingly 
show that muonic physics,
the physics of two-particle and three-particle muonic systems is currently an urgent problem that requires 
appropriate theoretical studies, calculations of observable quantities with high accuracy.

In the theoretical study of the energy levels of three-particle electron-muon-nucleus systems, two methods 
are usually used. One of them is the variational method, which allows to find wave functions and energies 
with very high accuracy \cite{huang,drachman,chen,af1,af2,korobov1,korobov2,korobov3,drake} (see other 
references in the review paper \cite {korobov3}).
Basically, theoretical studies were focused on muon-electron helium, since measurements of the hyperfine 
structure of the ground state were performed for it \cite{gladish,gardner}:
\begin{equation}
\label{eq2}
\Delta\nu^{hfs}_{exp}(\mu e^3_2He)=4166.3(2)~MHz,~~~
\Delta\nu^{hfs}_{exp}(\mu e^4_2He)=4465.004(29)~MHz.
\end{equation}

Analytical approach for calculating the energy levels of such three particle systems was formulated 
in \cite{lm1,lm2} and applied to calculate the hyperfine structure of the spectrum and the electronic Lamb shift 
in \cite{borie,amusia,km1,km2,sgk,apm1}. It is based on the use of the perturbation theory (PT) method with respect 
to two small parameters: the fine structure constant $\alpha$ and the electron - muon mass ratio. This approach 
has certain advantages, like any other analytical method, but in order to achieve high calculation accuracy, 
it is necessary to calculate numerous corrections in higher orders of perturbation theory.

In our previous paper \cite{apm1} we calculated the electron Lamb shift $(2P-2S)$ and the energy interval 
$(2S-1S)$ using the analytical method in muon-electron ions of lithium, beryllium and boron. We showed 
that the total value of the electronic Lamb shift strongly depends on the charge of the nucleus, so that 
in the transition from the lithium nucleus to the boron nucleus the magnitude of the shift undergoes 
a sharp decrease in the case of the beryllium nucleus. In this paper, we continue to study \cite{apm1} 
the energy levels of muon-electron ions of lithium, beryllium, and boron in the hyperfine part 
of the energy spectrum.

\section{Method for calculating basic contributions to HFS}\label{sec2} 

The Coulomb interaction in three particle muon - electronic ions of lithium, beryllium and boron leads 
to the formation of bound states The main features of such three particle systems are:
\begin{enumerate}
\item The lifetime of such systems is determined by the muon lifetime $\tau_\mu=2.1969811(22)\cdot 10^{-6}$ s. 
During this time, the muon manages to make about $10^{13}$ rotations around the nucleus.
\item The particle masses satisfy the inequality $m_e \ll m_\mu \ll M$, where $m_e$ is the electron mass, 
$m_\mu$ is the muon mass, $M$ is the nucleus mass. This leads to the fact that the muon is about 200 times 
closer to the nucleus than an electron. We can assume that the electron moves in the field of the quasinucleus, 
which is formed by the muon and the nucleus.
\item The hyperfine structure of the energy spectrum in the ground state arises from the interaction of particle 
spins: ${\bf s}_e$ is the electron spin, ${\bf s}_\mu$ is the muon spin, ${\bf I}$ is the nucleus spin. 
We consider as nuclei of lithium, beryllium and boron isotopes with nuclear spin $I=3/2$.
\end{enumerate}

To calculate the energy levels by the analytical perturbation theory method, we divide the Hamiltonian 
of the system into several parts, separating the main contribution of the Coulomb interaction $H_0$ in the form:
\begin{equation}
\label{eq3}
H=H_0+\Delta H+\Delta H_{rec}+\Delta H_{vp}+\Delta H_{str}+\Delta H_{vert},~
H_0=-\frac{1}{2M_\mu}\nabla^2_\mu-\frac{1}{2M_e}
\nabla^2_e-\frac{Z\alpha}{x_\mu}-\frac{(Z-1)\alpha}{x_e},
\end{equation}
\begin{equation}
\label{eq4}
\Delta H=\frac{\alpha}{|{\bf x}_{\mu}-{\bf x}_{e}|}-\frac{\alpha}{x_e},~~~
\Delta H_{rec}=-\frac{1}{M}{\boldsymbol\nabla}_\mu\cdot{\boldsymbol\nabla}_e,
\end{equation}
where ${\bf x_\mu}$ and ${\bf x_e}$ are the radius vectors of the muon and electron relative to the nucleus, 
$Ze$ is the nucleus charge.
The terms $\Delta H_{vp}$, $\Delta H_{str}$ and $\Delta H_{vert}$ denote contributions of vacuum polarization 
effects, effects of nucleus structure and vertex corrections. The reduced masses in the muon-nucleus 
and electron-nucleus subsystems are equal to
\begin{equation}
\label{eq5}
M_e=\frac{m_eM}{(m_e+M)},~~~M_\mu=\frac{m_\mu M}{(m_\mu+M)}.
\end{equation}

In the initial approximation, which is determined by the Hamiltonian $H_0$, the wave function 
of the system has a simple analytical form
\begin{equation}
\label{eq6}
\Psi_0({\bf x_e},{\bf x_\mu})=\psi_{e0}({\bf x_e})\psi_{\mu 0}({\bf
x_\mu})=\frac{1}{\pi} (W_e W_\mu)^{3/2}e^{-W_\mu x_\mu}e^{-W_e x_e},~
W_\mu =Z\alpha M_\mu,~W_e=(Z-1)\alpha M_e,
\end{equation}
which allows to accurately calculate the corrections using the perturbation theory. 
The Hamiltonian of the hyperfine interaction in the case of ground state can be presented in the form:
\begin{equation}
\label{eq7}
\Delta H^{hfs}=\tilde a({\bf S}_\mu\cdot{\bf I})-
\tilde b({\bf S}_e\cdot{\bf S}_\mu)+
\tilde c({\bf S}_e\cdot{\bf I}),
\end{equation}
where the coefficient functions $\tilde a$, $\tilde b$ and $\tilde c$  are presented in the form 
of the expansion by the perturbation theory. 
In leading order, these functions have the form:
\begin{equation}
\label{eq8}
\tilde a_0=\frac{2\pi\alpha}{3}\frac{g_Ng_\mu}{m_pm_\mu} \delta({\bf x}_\mu),~
\tilde b_0=\frac{2\pi\alpha}{3}\frac{g_\mu g_e}{m_\mu m_e}\delta({\bf x}_\mu-{\bf x}_e),~
\tilde c_0=\frac{2\pi\alpha}{3}\frac{g_eg_N}{m_em_p}\delta({\bf x}_e),
\end{equation}
where $g_e=2(1+a_e)$, $g_\mu=2(1+a_\mu)$ and $g_N=\frac{\mu_N}{I}$ are gyromagnetic factors of the electron, muon 
and nucleus, $\mu_N$ is a nucleus magnetic moment, $a_{e,\mu}$ are anomalous magnetic moments of the electron 
and muon.

Averaging the Hamiltonian \eqref{eq7} over the wave functions of the ground state, we obtain:
\begin{equation}
\label{eq9}
\nu=\langle \Delta H^{hfs}_0\rangle=a\ \langle {\bf I}\cdot{\bf S}_\mu\rangle-b\ \langle {\bf
S}_\mu\cdot{\bf S}_e\rangle +c\ \langle {\bf S}_e\cdot{\bf I}\rangle,
\end{equation}
where the coefficients 
\begin{equation}
\label{eq10}
a=\sum_{i=0}^\infty a_i,~~~ b=\sum_{i=0}^\infty b_i, ~~~c=\sum_{i=0}^\infty c_i
\end{equation}
are determined by different matrix elements by the perturbation theory. 
Using \eqref{eq6}, in leading order, we obtain the following contributions to $a$, $b$, $c$ 
(below, the values 
for lithium, beryllium, and boron nuclei are given line by line, and for helium nuclei, see Table~\ref{tb1}):
\begin{equation}
\label{eq10a}
a_0=\frac{g_N g_\mu}{4}\frac{m_e}{m_p}\left(\frac{W_\mu}{W_e}\right)^3\nu_F=
\begin{cases}
6.08349\cdot 10^8~MHz\\
-5.26948\cdot 10^8~MHz\\
23.66123\cdot 10^8~MHz.
\end{cases},~
\nu_F=\frac{8\alpha W_e^3}{3m_e m_\mu},
\end{equation}
\begin{equation}
\label{eq11}
b_0=\nu_F\frac{g_eg_\mu}{4}\frac{1}{\left(1+\frac{W_e}{W_\mu}\right)^3}=
\begin{cases}
35830.5299~MHz\\
120791.0324~MHz\\
286127.0374~MHz
\end{cases},
\end{equation}
\begin{equation}
\label{eq12}
c_0=\nu_F\frac{m_\mu}{m_p}\frac{g_eg_N}{4}=
\begin{cases}
4422.8997~MHz\\
-5397.5666~MHz\\
29216.4109~MHz.
\end{cases}.
\end{equation}

\begin{figure}
\centering
\includegraphics[scale=0.5]{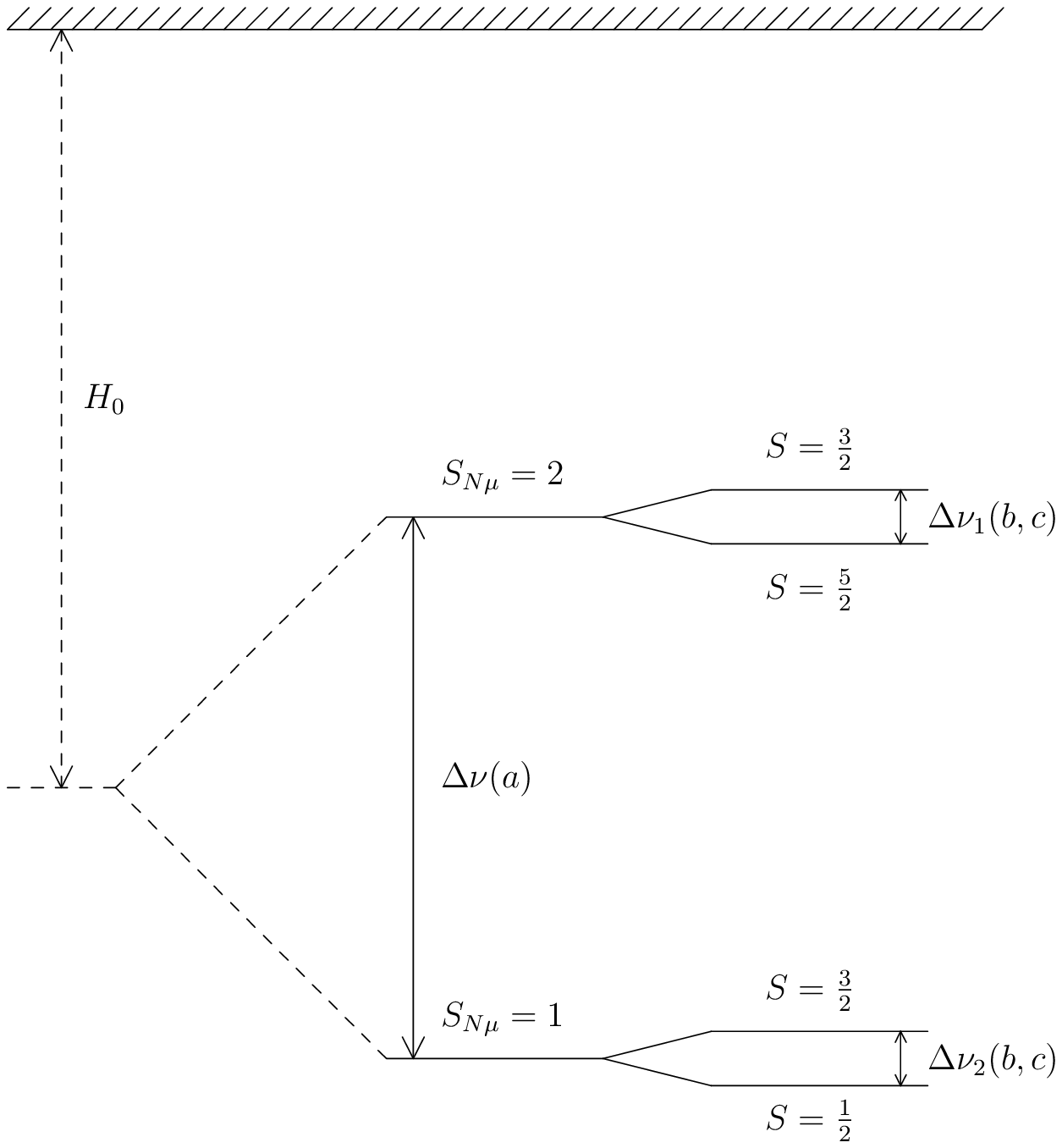}
\caption{Hyperfine splitting of the ground state of muon-electron ions of lithium, beryllium and boron.}
\label{fig_libeb}
\end{figure}

For the calculation of matrix elements from the product of spin operators we use the following transformation 
of basis wave functions \cite{iis}:
\begin{equation}
\label{eq13}
\Psi_{S_{N\mu}SS_z}=\sum_{S_{Ne}} (-1)^{S_\mu+I+S_e+S}\sqrt{(2S_{N\mu}+1)(2S_{Ne}+1)}
\left\{\begin{array}{ccc}
S_e& S_N & S_{Ne} \\
S_\mu & S & S_{N\mu} \\
\end{array}\right\}\Psi_{S_{Ne}SS_z},
\end{equation}
where $S_{N\mu}$ is spin of muon - nucleus subsystem, $S_{Ne}$ is spin of electron - nucleus subsystem, $S$ 
is total spin of three particle system. The properties of $6j$ - symbols are discussed in \cite{iis}.

The given numerical values of the main contributions to the coefficients $a$, $b$, $c$ show that the system 
has small intervals of the hyperfine structure, which are determined by the quantities $b$ and $c$. An approximate 
scheme of energy level splitting due to hyperfine interaction of the ground state is shown in Fig.~\ref{fig_libeb}. 
The coefficient $b_0$ was calculated using the $g$-factor of the electron $g_e\approx 2$. The correction connected 
with the anomalous magnetic moment of an electron in this interaction is considered in section \ref{sec5}. 
When calculating other coefficients, the following values of gyromagnetic factors are used:
$g_e=2(1+\kappa_e)=2(1+ 1.15965218111(74)\cdot 10^{-3})$,
$g_\mu=2(1+\kappa_\mu)=2\cdot (1+1.16592069(60)\cdot 10^{-3})$,
$g_N(^7_3Li)=2.170951$, $g_N(^9_4Be)=-0.784955$, $g_N(^{11}_5B)=1.792433$ \cite{stone}.

The average value of the Hamiltonian of hyperfine interaction $\Delta H_0^{hfs}$ calculated in 
$\psi_{S_{N\mu}SS_z}$ basis has the form:
\begin{equation}
\label{eq14}
\nu =\left( 
\begin{array}{ccccc}
\  & {\mathrm{\Psi }}_{1,\frac{1}{2},S_z} & {\mathrm{\Psi }}_{1,\frac{3}{2},S_z} & {\mathrm{\Psi }}_{2,\frac{3}{2},S_z} & {\mathrm{\Psi }}_{2,\frac{5}{2},S_z} \\ 
{\mathrm{\Psi }}_{1,\frac{1}{2},S_z} & -\frac{5}{4}a-\frac{1}{4}b-\frac{5}{4}c & 0 & 0 & 0 \\ 
{\mathrm{\Psi }}_{1,\frac{3}{2},S_z} & 0 & -\frac{5}{4}a+\frac{1}{8}b+\frac{5}{8}c & -\frac{\sqrt{15}}{8}b+
\frac{\sqrt{15}}{8}c & 0 \\ 
{\mathrm{\Psi }}_{2,\frac{3}{2},S_z} & 0 & -\frac{\sqrt{15}}{8}b+\frac{\sqrt{15}}{8}c & \frac{3}{4}a+\frac{3}{8}b-
\frac{9}{8}c & 0 \\ 
{\mathrm{\Psi }}_{2,\frac{5}{2},S_z} & 0 & 0 & 0 & \frac{3}{4}a-\frac{1}{4}b+\frac{3}{4}c 
\end{array}
\right).
\end{equation}

After diagonalizing this matrix, we get four eigenvalues that define the hyperfine structure:
\begin{equation}
\label{eq14a}
\begin{array}{c}
\nu_1 \left(S_{N\mu }=\frac{1}{2},\ \ \ S=1\right)=-\frac{5}{4}a-\frac{1}{4}b-\frac{5}{4}c, \\
\nu_2 \left(S_{N\mu }=\frac{3}{2},\ \ \ S=1\right)=\frac{1}{4}\left(-a+b-c-\sqrt{16a^2+4b^2+16c^2+4ab-28ac-11bc}\right),\\
\nu_3 \left(S_{N\mu }=\frac{3}{2},\ \ \ S=2\right)=\frac{1}{4}\left(-a+b-c+\sqrt{16a^2+4b^2+16c^2+4ab-28ac-11bc}\right),\\
\nu_4 \left(S_{N\mu }=\frac{5}{2},\ \ \ S=2\right)=\frac{3}{4}a-\frac{1}{4}b+\frac{3}{4}c.
\end{array}
\end{equation}

As long as $a\gg b$ and $a\gg c$ we can use expansions in $b/a$, $c/a$ and represent small intervals 
of the hyperfine structure in the form:
\begin{equation}
\label{eq15}
\Delta\nu_1^{hfs}=\nu_3-\nu_4=\frac{5(b-3c)}{8}+O\left(\frac{b}{a},\frac{c}{a}\right),~
\Delta\nu_2^{hfs}=\nu_2-\nu_1=\frac{3(b+5c)}{8}+O\left(\frac{b}{a},\frac{c}{a}\right).
\end{equation}

As it follows from \eqref{eq11}, $b_0$ contains the recoil effects over $W_e/W_\mu$ in the leading order in $\alpha$. 
The same recoil effects also occur in the second order of PT in $\Delta H$.

\section{Recoil corrections in second order of perturbation theory}\label{sec3} 

Recoil corrections of order $\alpha^4\frac{W_e}{W_\mu}$, $\alpha^4\frac{W^2_e}
{W^2_\mu}\ln\frac{W_e}{W_\mu}$ and $\alpha^4\frac{W^2_e}{W^2_\mu}$ occur in the second order of perturbation theory. 
The contribution to the $b$ coefficient is determined by the following expression:
\begin{equation}
\label{eq15a}
b_1=2\int\Psi^\ast({\bf x}_e,{\bf x}_\mu)\tilde b_0({\bf x}_e-{\bf x}_\mu)
\tilde G({\bf x}_e,{\bf x}_\mu;{\bf x'}_e,{\bf x'}_\mu)
\Delta H({\bf x'}_e,{\bf x'}_\mu)\Psi({\bf x'}_e,{\bf x'}_\mu)d{\bf x}_e
d{\bf x}_\mu d{\bf x'}_ed{\bf x'}_\mu,
\end{equation}
where the reduced Coulomb Green's function has the form:
\begin{equation}
\label{eq16}
\tilde G({\bf x}_e,{\bf x}_\mu;{\bf x'}_e,{\bf x'}_\mu)=\sum_{n,n'\not =0}
\frac{\psi_{\mu n}({\bf x}_\mu)
\psi_{en'}({\bf x}_e)\psi^\ast_{\mu n}({\bf x'}_\mu)\psi^\ast_{en'}({\bf x'}_e)}{E_{\mu 0}+
E_{e0}-E_{\mu n}-E_{en'}}.
\end{equation}

It is convenient to divide the sum over muon states in \eqref{eq16} into two parts with $n=0$ and $n\not=0$. 
For the first part we get:
\begin{equation}
\label{eq17}
b_1(n=0)=\frac{4\pi \alpha}{3}\frac{g_eg_\mu}{m_em_\mu}\int|\psi_{\mu 0}({\bf x}_3)|^2
\psi^\ast_{e0}({\bf x}_3)\sum_{n'\not=0}^\infty
\frac{\psi_{en'}({\bf x}_3)\psi^\ast_{en'}({\bf x}_1)}{E_{e0}-E_{en'}}V_\mu({\bf x}_1)
\psi_{e0}({\bf x}_1)d{\bf x}_1d{\bf x}_3,
\end{equation}
\begin{equation}
\label{eq18}
V_\mu({\bf x}_1)=\int\psi^\ast_{\mu 0}({\bf x}_2)\left[\frac{\alpha}{|{\bf x}_2-{\bf x}_1|}-\frac{\alpha}{x_1}\right]\psi_{\mu 0}({\bf x}_2)d{\bf x}_2=
-\frac{\alpha}{x_1}(1+W_\mu x_1)e^{-2W_\mu x_1}.
\end{equation}
The reduced Coulomb Green's function of an electron in \eqref{eq17} is determined by \cite{hameka}:
\begin{equation}
\label{eq19}
G_e({\bf x}_1,{\bf x}_3)=\sum_{n\not =0}^\infty\frac{\psi_{en}({\bf
x}_3) \psi_{en}^\ast({\bf x}_1)}{E_{e0}-E_{en}}=-\frac{W_e M_e}{\pi}e^{-W_e(x_1+x_3)}
\Biggl[\frac{1}{2W_e x_>}-
\end{equation}
\begin{displaymath}
-\ln(2W_e x_>)-\ln(2W_e x_<)+Ei(2W_e x_<)+
\frac{7}{2}-2C-W_e(x_1+x_3)+\frac{1-e^{2W_e x_<}}{2W_e x_<}\Biggr],
\end{displaymath}
where $x_<=\min(x_1,x_3)$, $x_>=\max(x_1,x_3)$, $C=0.577216\ldots$ is the Euler constant and $Ei(x)$ is integral 
exponential function. Then the coordinate integration in \eqref{eq17} can be performed analytically, and 
the obtained result can be represented as an expansion in $W_e/W_\mu$:
\begin{equation}
\label{eq20}
b_1(n=0)=\nu_F\frac{(1+\kappa_\mu)}{(Z-1)}\left[\frac{11}{8}\frac{W_e}{W_\mu}-\frac{1}{16}
\frac{W_e^2}{W_\mu^2}\left(64\ln\frac{W_e}{W_\mu}+64\ln 2+7\right)\right].
\end{equation}
Exited states of muon $(n\not=0)$ give second part of contribution to $b$ coefficient:
\begin{equation}
\label{eq21}
b_1(n\not=0)=\frac{4\pi\alpha}{3}\frac{g_eg_\mu}{m_em_\mu}\int\psi^\ast_{\mu 0}({\bf x}_3)
\psi^\ast_{e 0}({\bf x}_3)
\sum_{n\not=0}\psi_{\mu n}({\bf x}_3)\psi^\ast_{\mu n}({\bf x}_2)G_e({\bf x}_3,{\bf x}_1,z)\times
\end{equation}
\begin{displaymath}
\left[\frac{\alpha}{|{\bf x}_2-{\bf x}_1|}-\frac{\alpha}{x_1}\right]\psi_{\mu 0}({\bf x}_2)
\psi_{e 0}({\bf x}_1)d{\bf x}_1 d{\bf x}_2d{\bf x}_3.
\end{displaymath}
where we introduce the electron Green's function
\begin{equation}
\label{eq22}
G_e({\bf x}_3,{\bf x}_1,z)=\sum_{n'=0}^\infty\frac{\psi_{en'}({\bf x}_3)
\psi^\ast_{en'}({\bf x}_1)}{z-E_{en'}}=
\sum_{n'=0}^\infty\frac{\psi_{en'}({\bf x}_3)\psi^\ast_{en'}({\bf x}_1)}
{E_{\mu 0}+E_{e0}-E_{\mu n}-E_{en'}}.
\end{equation}
The term $(-\alpha/x_1)$ in \eqref{eq21} does not contribute due to the orthogonality of the muon wave functions. 
To perform further analytic integration in \eqref{eq19}, we replace $G_e$ approximately with the free Green's 
function \cite{lm1,lm2}:
\begin{equation}
\label{eq23}
G_e({\bf x}_3,{\bf x}_1,E_{\mu 0}+E_{e0}-E_{\mu n})\to G_{e0}({\bf x}_3-{\bf x}_1,E_{\mu 0}+E_{e0}-E_{\mu n})=-\frac{M_e}{2\pi}\frac{e^{-\beta|{\bf x}_3-{\bf x}_1|}}{|{\bf x}_3-{\bf x}_1|},
\end{equation}
where $\beta=\sqrt{2M_e(E_{\mu n}-E_{e0}-E_{\mu 0})}$. In addition, we approximately replace wave functions of an electron 
in \eqref{eq19} by their values at zero $\psi_{e0}(0)$. The terms omitted in this approximation 
can give a second-order contribution with respect to $\frac{W_e}{W_\mu}$ in $b$. The results of numerical 
integration in \cite{lm1} with the exact Green's function of the electron in the case of muonic helium show 
that the terms used in the \eqref{eq23} approximation are numerically small.

After these approximations, the integration over the ${\bf x}_1$ coordinate gives the following result:
\begin{equation}
\label{eq24}
\int\frac{e^{-\beta|{\bf x}_3-{\bf x}_1|}}{|{\bf x}_3-{\bf x}_1|}
\frac{d{\bf x}_1}{|{\bf x}_2-{\bf x}_1|}=
4\pi\left[\frac{1}{\beta}-\frac{1}{2}|{\bf x}_3-{\bf x}_2|+\frac{1}{6}\beta|{\bf x}_3-{\bf x}_2|^2-
\frac{\beta^2}{24}|{\bf x}_3-{\bf x}_2|^3+\ldots\right],
\end{equation}
where an expansion of $e^{-\beta|{\bf x}_2-{\bf x}_3|}$ in $\beta|{\bf x}_2-{\bf x}_3|$ is done. 
This expansion is equivalent to the expansion in powers of $\sqrt{W_e/W_\mu}$. The first expansion 
term $\beta^{-1}$ in \eqref{eq24} does not contribute to \eqref{eq21}. The second expansion term 
in \eqref{eq24} gives the leading order contribution in $\sqrt{W_e/W_\mu}$: $-\nu_F\frac{35W_e}{8(Z-1)W_\mu}$. 
To increase the accuracy of the result, we also consider the third term on the right side \eqref{eq24}, 
which leads to the following integral:
\begin{equation}
\label{eq25}
\int \psi^\ast_{\mu 0}({\bf x}_3)\sum_{n}\sqrt{2M_e(E_{\mu n}-E_{\mu 0})}
\psi_{\mu n}({\bf x}_3)\psi^\ast_{\mu n}({\bf x}_2)({\bf x}_2\cdot{\bf x}_3)
\psi_{\mu 0}({\bf x}_2)d{\bf x}_2d{\bf x}_3=
\sqrt{\frac{W_e Z}{W_\mu^3(Z-1)}}S_{\frac{1}{2}},
\end{equation}
where we introduce the quantity
\begin{equation}
\label{eq26}
S_{1/2}=\sum_{n}\left(\frac{E_{\mu n}-E_{\mu 0}}{R_\mu}\right)^{1/2}|\langle\mu 0|
\frac{\bf x}{a_\mu}|\mu n\rangle|^2, ~~~R_\mu=\frac{1}{2}M_\mu(Z\alpha)^2.
\end{equation}

A contribution to \eqref{eq26} comes from matrix elements for discrete and continuous states, which are presented 
in \cite{bs}. The numerical contributions of discrete and continuous states to \eqref{eq26} have the form:
\begin{equation}
\label{eq27}
S_{\frac{1}{2}}^{d}=\sum_{n}\frac{2^{8}n^6(n-1)^{2n-\frac{9}{2}}}{(n+1)^{2n+\frac{9}{2}}}=1.90695...,
\end{equation}
\begin{equation}
S_{\frac{1}{2}}^{c}=\int_0^\infty \frac{2^8kdk}{(k^2+1)^{9/2}(1-e^{-\frac{2\pi}{k}})}
\left|\left(\frac{1+ik}{1-ik}\right)^{i/k}\right|=1.03111...  .
\end{equation}

Adding the recoil corrections in the second order of perturbation theory, we get the total recoil correction 
to $b$ of order $\alpha^4$ as follows:
\begin{equation}
\label{eq28}
b_{rec}=\nu_F\frac{(1+\kappa_\mu)}{(Z-1)}\bigl[-3\frac{W_e}{W_\mu}+
\frac{231W_e^2}{32W_\mu^2}-
\frac{4W_e^2}{W_\mu^2}\ln\frac{2W_e}{W_\mu}+\frac{4W_e}{3W_\mu}\sqrt{\frac{W_e Z}{W_\mu
(Z-1)}}S_{1/2}\bigr].
\end{equation}

In the second order of perturbation theory we have the same contribution in coefficient $c$. To calculate it, 
it is necessary to choose the hyperfine part of the perturbation operator in the form 
$\Delta H^{hfs}_0({\bf x}_e)=\frac{2\pi\alpha}{3} \frac{g_eg_N}{m_em_p}\ delta({\bf x}_e)$ in a general 
expression like \eqref{eq21}. Using then the obvious simplifications connected with the $\delta$-function, 
one can transform the recoil correction to $c$ as follows:
\begin{equation}
\label{eq29}
c_1=\frac{4\pi\alpha}{3}\frac{g_eg_N}{m_em_p}\int \psi^\ast_{e0}(0)
\tilde G_e(0,{\bf x}_1)V_\mu({\bf x}_1)\psi_{e0}({\bf x}_1)d{\bf x}_1.
\end{equation}
The reduced Coulomb Green's function of an electron with one zero argument in this equation is equal to
\begin{equation}
\label{eq30}
\tilde G_e(0,{\bf x})=\sum_{n\not =0}^\infty\frac{\psi_{en}(0)
\psi^\ast_{en}({\bf x})}{E_{e0}-E_{en}}=-\frac{W_eM_e}{\pi}e^{-W_e x}
\left[\frac{1}{2W_e x}-\ln 2W_e x+\frac{5}{2}-C-W_e x\right].
\end{equation}
As a result of analytical calculation of matrix elements over coordinate variables we get a contribution 
to the coefficient $c$. It can be presented in the form of an expansion in $W_e/W_\mu$:
\begin{equation}
\label{eq31}
c_1=c_0\frac{2}{(Z-1)}\left[\frac{3W_e}{2W_\mu}+2\frac{W_e^2}{W_\mu^2}\left(\frac{1}{4}-
\ln\frac{W_e}{W_\mu}\right)\right].
\end{equation}
Numerical values of contribution \eqref{eq31} for various muon - electron ions are presented in Table~\ref{tb1}.

\section{Effects of vacuum polarization}\label{sec4}

\begin{figure}[htbp]
\centering
\includegraphics[scale=0.7]{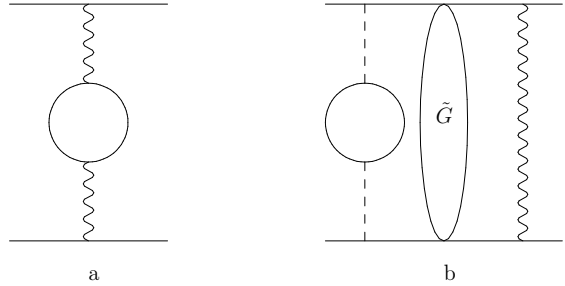}
\caption{Correction of vacuum polarization. Dashed line denotes the contribution of the Coulomb photon. 
The wavy line denotes the hyperfine part of the Breit potential. $\tilde G$ denotes the reduced Coulomb 
Green's function.}
\label{fig2}
\end{figure}

\begin{figure}[htbp]
\centering
\includegraphics[scale=0.7]{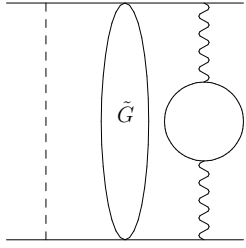}
\caption{Effects of vacuum polarization in second order of perturbation theory. Dashed line denotes the potential 
$\Delta H$ (3). The wavy line denotes the hyperfine part of the Breit potential.}
\label{fig3}
\end{figure}

Among other corrections in the energy spectrum of muonic atoms and ions, corrections of vacuum polarization \cite{apm2016,apm2017} stand out. The one-loop vacuum polarization, which is taken into account in this paper, 
gives a fifth-order contribution in $\alpha$ to the HFS. The corresponding interaction amplitudes in the first 
and second orders of the perturbation theory are shown schematically in Fig.~\ref{fig2}-\ref{fig3}.

The correction for vacuum polarization in the first order of PT is related to the modification of the hyperfine 
part of the Hamiltonian \eqref{eq7} (see Fig.~\ref{fig2}(a)), which has the form in the case of muon-electron 
and electron-nuclear interactions:
\begin{equation}
\label{eq32}
\Delta V_{vp,e\mu}^{hfs}({\bf x}_{e\mu})=-\frac{2\alpha g_e g_\mu}{3m_em_\mu}
({\bf S}_e\cdot {\bf S}_\mu)\frac{\alpha}{3\pi}\int_1^\infty
\rho(\xi)d\xi\left[\pi\delta({\bf
x_{e\mu}})-\frac{m_e^2\xi^2}{x_{e\mu}}e^{-2m_e\xi x_{e\mu}} \right],
\end{equation}
\begin{equation}
\label{eq33}
\Delta V_{vp,eN}^{hfs}({\bf x}_e)=\frac{2\alpha g_e g_N}{3m_em_p} 
({\bf S}_e\cdot {\bf I})\frac{\alpha}{3\pi}\int_1^\infty
\rho(\xi)d\xi\left[\pi\delta({\bf x_e})-\frac{m_e^2\xi^2}{x_e}e^{-2m_e\xi x_e} \right],
\end{equation}
\begin{equation}
\label{eq33a}
\rho(\xi)=\frac{\sqrt{\xi^2-1}(2\xi^2+1)}{\xi^4}.
\end{equation}

Matrix element of the potential \eqref{eq32} with wave functions \eqref{eq6} gives the contribution 
to coefficient $b$:
\begin{equation}
\label{eq34}
b_{vp}=\frac{8\alpha^2}{9m_em_\mu}\frac{W_e^3W_\mu^3}{\pi^3}\int_1^\infty\rho(\xi)d\xi
\int d{\bf x}_e\int d{\bf x}_\mu e^{-2W_\mu x_\mu}e^{-2W_e x_e}\times
\end{equation}
\begin{displaymath}
\times\left[\pi\delta({\bf x_\mu}-{\bf x}_e)-\frac{m_e^2\xi^2}{|{\bf x}_\mu-{\bf x}_e|}
e^{-2m_e\xi|{\bf x}_\mu-{\bf x}_e|}\right].
\end{displaymath}
Both integrals over coordinates of muon and electron in \eqref{eq34} can be calculated analytically:
\begin{equation}
\label{eq35}
I_1=\int d{\bf x}_e\int d{\bf x}_\mu e^{-2W_\mu x_\mu}e^{-2W_e x_e}\pi
\delta({\bf x_\mu}-{\bf x}_e)
=\frac{\pi^2}{W_\mu^3\left(1+\frac{W_e}{W_\mu}\right)^3},
\end{equation}
\begin{equation}
\label{eq36}
I_2=\int d{\bf x}_e\int d{\bf x}_\mu e^{-2W_\mu x_\mu}e^{-2W_e x_e}\frac{m_e^2\xi^2}
{|{\bf x}_\mu-{\bf x}_e|}e^{-2m_e\xi|{\bf x}_\mu-{\bf x}_e|}=
\end{equation}
\begin{displaymath}
=\frac{\pi^2m_e^2\xi^2}{W_\mu^5}\frac{\left[\frac{W_e^2}{W_\mu^2}+\left(1+
\frac{m_e\xi}{W_\mu}\right)^2+\frac{W_e}{W_\mu}\left(3+\frac{2m_e\xi}{W_\mu}\right)
\right]}{\left(1+\frac{W_e}{W_\mu}\right)^3\left(1+\frac{m_e\xi}
{W_\mu}\right)^2\left(\frac{W_e}{W_\mu}+\frac{m_e\xi}{W_\mu}\right)^2}.
\end{displaymath}
Separately integrals over spectral parameter $\xi$ in \eqref{eq35} and \eqref{eq36} are divergent. 
But their sum is finite and can be presented as follows:
\begin{equation}
\label{eq37}
b_{vp}=\nu_F\frac{\alpha W_e}{3\pi W_\mu\left(1+\frac{W_e}{W_\mu}\right)^3}
\int_1^\infty\rho(\xi)d\xi
\frac{\left[\frac{W_e}{W_\mu}+2\frac{m_e\xi}{W_\mu}\frac{W_e}{W_\mu}+
\frac{m_e\xi}{W_\mu}\left(2+\frac{m_e\xi}{W_\mu}\right)\right]}
{\left(1+\frac{m_e\xi}{W_\mu}\right)^2\left(\frac{W_e}{W_\mu}+
\frac{m_e\xi}{W_\mu}\right)^2}.
\end{equation}

The order of contribution \eqref{eq37} is determined by two small parameters $\alpha$ and $W_e/W_\mu$. 
The correction $b_{vp}$ has the fifth order in $\alpha$ and the first order in $W_e/W_\mu$. The integration over 
$\xi$ in \eqref{eq37} is done numerically. The result is presented in Table~\ref{tb1}.

The contribution of the muon vacuum polarization is much smaller than \eqref{eq37} and is not taken into account 
when obtaining the total numerical value of the hyperfine splitting. We also neglect the contribution 
of two-loop vacuum polarization, which is suppressed by an additional factor $\alpha/\pi$.

The contribution of correction of one - loop vacuum polarization to the coefficient $c$ has the order $\alpha^6$. 
It can be calculated by the same way using potential \eqref{eq33} ($\alpha_1=W_e/m_e$). After integration 
over all variables, including $\xi$, we obtain:
\begin{equation}
\label{eq38}
c_{vp}=\nu_F\frac{\alpha g_N m_\mu}{6\pi m_p}\frac{\sqrt{1-\alpha_1^2}
(6\alpha_1+\alpha_1^3-3\pi)+(6-3\alpha_1^2+6\alpha_1^4)\arccos\alpha_1}{3\alpha_1^3
\sqrt{1-\alpha_1^2}}.
\end{equation}

When calculating corrections in the second order of PT, it is necessary to use the following expressions 
for the Coulomb potentials as one of the perturbation operators, taking into account the vacuum polarization 
effect \cite{apm2015,egs,t4}:
\begin{equation}
\label{eq39}
\Delta V_{vp}^{eN}(x_e)=\frac{\alpha}{3\pi}\int_1^\infty
\rho(\xi)\left(-\frac{Z\alpha}{x_e}\right) e^{-2m_e\xi
x_e}d\xi,
\end{equation}
\begin{equation}
\label{eq40}
\Delta V_{vp}^{\mu N}(x_\mu)=\frac{\alpha}{3\pi}\int_1^\infty
\rho(\xi)\left(-\frac{Z\alpha}{x_\mu}\right) e^{-2m_e\xi x_\mu}d\xi,
\end{equation}
\begin{equation}
\label{eq41}
\Delta V_{vp}^{e\mu}(|{\bf x}_e-{\bf
x}_\mu|)=\frac{\alpha}{3\pi}\int_1^\infty
\rho(\xi)\frac{\alpha}{x_{e\mu}} e^{-2m_e\xi x_{e\mu}}d\xi,
\end{equation}
where $x_{e\mu}=|{\bf x}_e-{\bf x}_\mu|$. 

The original integral expression for the contribution to $b$ from the electron-nuclear potential \eqref{eq39} 
in second order of PT has the form:
\begin{equation}
\label{eq42}
b_{vp,~sopt}^{eN}=\frac{4\pi\alpha g_eg_\mu}{3m_em_\mu}\int 
d{\bf x}_1\int d{\bf x}_2\int d{\bf x}_3\frac{\alpha}{3\pi}\int_1^\infty
\rho(\xi)d\xi\psi^\ast_{\mu 0}({\bf x}_3)\psi^\ast_{e 0}({\bf x}_3)\times
\end{equation}
\begin{displaymath}
\times\sum_{n,n'\not =0}^\infty\frac{\psi_{\mu n}({\bf x}_3) 
\psi_{e n'}({\bf x}_3)\psi^\ast_{\mu n}({\bf x}_2)\psi^\ast_{e n'}({\bf x}_1)}
{E_{\mu 0}+E_{e0}-E_{\mu n}-E_{en'}}\left(-\frac{Z\alpha}{x_1}\right)
e^{-2m_e\xi x_1}\psi_{\mu 0}({\bf x}_2) \psi_{e0}({\bf x}_1),
\end{displaymath}
where the index "sopt" is used
to denote the second-order of PT contribution. The summation in \eqref{eq42} is performed 
over the entire set of electron and muon states, excluding the state with $n,n'=0$. Using the orthogonality 
of muon wave functions, the correction \eqref{eq42} can be represented in integral form:
\begin{equation}
\label{eq43}
b_{vp,~sopt}^{eN}=\nu_F\frac{Z\alpha a_2^2}{3(Z-1)\pi}\int_1^\infty \rho(\xi)d\xi
\int _0^\infty x_3^2 dx_3\int_0^\infty x_1 dx_1
e^{-\gamma_1\left(1+\gamma_3\right)x_1}e^{-x_3\left(1+\gamma_1\right)}\times
\end{equation}
\begin{displaymath}
\Bigl[\frac{1}{\gamma_1 x_>}-\ln\left(\gamma_1x_<\right)-\ln
\left(\gamma_1x_>\right)+Ei\left(\gamma_1 x_<\right)+
\frac{7}{2}-2C-\frac{\gamma_1}{2}(x_1+x_3)+
\end{displaymath}
\begin{displaymath}
+\frac{1-e^{\gamma_1 x_<}}{\gamma_1 x_<}\Bigr]=
\nu_F\frac{Z\alpha}{3(Z-1)\pi(1+\gamma_1)^4}\int_1^\infty \frac{\rho(\xi)d\xi}{(1+\gamma_3)^3
(1+\gamma_3\gamma_1+\gamma_1)^2}
\end{displaymath}
\begin{displaymath}
\bigl[
3+\gamma_3(7+2\gamma_3)+\gamma_1(6+\gamma_3(20+\gamma_3(13+\gamma_3)))+(1+\gamma_3)
(3+2\gamma_3(5+2\gamma_3))\gamma_1^2+
\end{displaymath}
\begin{displaymath}
+2(1+\gamma_3)(1+\gamma_1)(1+\gamma_1+\gamma_3\gamma_1)\ln\bigl(1-\frac{\gamma_3}{(1+\gamma_3)
(1+\gamma_1)}\bigr)\bigr],~
\gamma_3=\frac{m_e\xi}{W_e},~\gamma_1=\frac{W_e}{W_\mu}.
\end{displaymath}

The integration over particle coordinates is carried out analytically. The integration over $\xi$ is done numerically. 
Numerical results are presented in Table~\ref{tb1}.

A similar calculation can be performed in the case of a potential with muon-nuclear vacuum polarization \eqref{eq40}. 
The electron remains in the $1S$ state, and the reduced Coulomb Green's function of the system is transformed into 
the muon Green's function. In this case, the correction to the coefficient $b$ can be represented as an integral:
\begin{equation}
\label{eq44}
b_{vp,sopt}^{\mu N}
=\nu_F\frac{\alpha}{3\pi}\int_1^\infty\frac{\rho(\xi)d\xi}{(\gamma_3+1)^3(\gamma_1+1)^4(\gamma_3+\gamma_1+1)^2}
\bigl[\gamma_1(\gamma_3^3(\gamma_1+4)+\gamma_3^2(\gamma_1(2\gamma_1+13)+14)+
\end{equation}
\begin{displaymath}
\gamma_3(\gamma_1+1)(7\gamma_1+13)+3(\gamma_1+1)^2)+
2(\gamma_3+1)(\gamma_1+1)(\gamma_3+\gamma_1+1)^2\ln\frac{(\gamma_3+1)(\gamma_1+1)}{(\gamma_3+\gamma_1+1)}
\bigr].
\end{displaymath}

The second-order of PT correction to $b$, which is determined by the potential \eqref{eq32}, turns out to be the most 
difficult to calculate. In this case, it is necessary to take into account intermediate excited states for both 
the electron and the muon. We break this contribution into two parts. The first part, in which the muon 
is in the $1S$ intermediate state has the form:
\begin{equation}
\label{eq45}
b_{vp,sopt}^{\mu e}(n=0)=\frac{256\alpha^2 W_e^3 W_\mu^3}{9\pi m_e m_\mu} 
\int_0^\infty x_3^2dx_3e^{-(W_e+2W_\mu)x_3}\times
\end{equation}
\begin{displaymath}
\times\int_0^\infty x_1^2 dx_1 e^{-W_ex_1}\int_1^\infty\rho(\xi)
d\xi\Delta V_{vp, \mu}(x_1)G_e(x_1,x_3),
\end{displaymath}
where function $V_{vp, \mu}(x_1)$ is
\begin{equation}
\label{eq46}
\Delta V_{vp, \mu}(x_1)=\frac{W_\mu^3}{\pi}\int d{\bf x}_2 e^{-2W_\mu x_2}
\frac{\alpha}{|{\bf x}_1-{\bf x}_2|}e^{-2m_e\xi|{\bf x}_1-{\bf x}_2|}=
\end{equation}
\begin{displaymath}
=\frac{\alpha W^3_\mu}{x_1(W_\mu^2-m_e^2\xi^2)^2}
\left[W_\mu\left(e^{-2m_e\xi x_1}-e^{-2W_\mu x_1}\right)+
x_1(m_e^2\xi^2-W_\mu^2)e^{-2W_\mu x_1}\right].
\end{displaymath}
Substituting \eqref{eq46} into \eqref{eq45} and integrating over particle coordinates, we get:
\begin{equation}
\label{eq47}
b_{vp,sopt}^{\mu e}(n=0)=-\nu_F\frac{\alpha \gamma_1}{3\pi(Z-1)(1+\gamma_1)^4}\int_1^\infty\frac{\rho(\xi)d\xi}
{(1-\gamma_3^2)^2}\Bigr[
-\frac{6(-1+\gamma_3^2)}{(1+\gamma_1)^3}+\frac{(-6+11\gamma_3^2)}{(1+\gamma_1)^2}-
\end{equation}
\begin{displaymath}
\frac{(1+7\gamma_3^2)}{(1+\gamma_1)}+\frac{(-1+\gamma_3^2)}{(2+\gamma_1)^3}+\frac{(3-4\gamma_3^2)}{(2+\gamma_1)^2}+
\frac{(-2+7\gamma_3^2)}{(2+\gamma_1)}+\frac{2(-1+\gamma_3)\gamma_3^2}{(\gamma_3+\gamma_1)^3}-
\frac{\gamma_3(-1+\gamma_3(5+\gamma_3))}{(\gamma_3+\gamma_1)^2}+
\end{displaymath}
\begin{displaymath}
\frac{3+5\gamma_3^2}{(\gamma_3+\gamma_1)}+\frac{\gamma_3^3}{(1+\gamma_3+\gamma_1)^2}-\frac{5\gamma_3^2}
{1+\gamma_3+\gamma_1}+\frac{2\gamma_1(-2+\gamma_3^2-\gamma_1)}{(1+\gamma_1)^2}
\ln\frac{1+\gamma_1}{2+\gamma_1}+\frac{2\gamma_1}{(\gamma_3+\gamma_1)^2}\ln\frac{\gamma_3+\gamma_1}
{1+\gamma_3+\gamma_1}
\end{displaymath}
\begin{displaymath}
+\frac{2\gamma_1}{(1+\gamma_1)^2(\gamma_3+\gamma_1)^2}\left(-(-1+\gamma_3)^2(1+\gamma_3^2+2\gamma_3(1+\gamma_1)+
\gamma_1(3+\gamma_1)\ln\frac{\gamma_1}{1+\gamma_1}\right)
\Bigl].
\end{displaymath}
The integration over the parameter $\xi$ is performed numerically.
The second part of the correction under consideration to $b$ can initially be represented as:
\begin{equation}
\label{eq48}
b_{vp,sopt}^{\mu e}(n\not=0)=-\frac{4\alpha^2}{9\pi}\frac{g_eg_\mu}{m_em_\mu}\int d{\bf x}_3\int d{\bf
x}_2\int_1^\infty\rho(\xi)d\xi \psi^\ast_{\mu 0}({\bf x}_3)\psi^\ast_{e 0}({\bf x}_3)\times
\end{equation}
\begin{displaymath}
\times\sum_{n\not =0}\psi_{\mu n}({\bf x}_3)\psi^\ast_{\mu n}({\bf
x}_2) \frac{M_e}{2\pi}\frac{e^{-{\beta}|{\bf x}_3-{\bf
x}_1|}}{|{\bf x}_3-{\bf x}_1|}\frac{\alpha}{|{\bf x}_2-{\bf
x}_1|}e^{-2m_e\xi|{\bf x}_2-{\bf x}_1|} \psi_{\mu 0}({\bf
x}_2)\psi_{e 0}({\bf x}_1),
\end{displaymath}
where, as before, the exact electron Coulomb Green's function is replaced by the free one. Replacing also 
the electronic wave functions with their values at zero, we thus neglect the same recoil corrections and can 
perform analytical integration over ${\bf x}_1$:
\begin{equation}
\label{eq49}
J=\int d{\bf x}_1\frac{e^{-{\beta}|{\bf x}_3-{\bf x}_1|}}{|{\bf
x}_3-{\bf x}_1|} \frac{e^{-2m_e\xi|{\bf x}_2-{\bf x}_1|}}{|{\bf
x}_2-{\bf x}_1|}= -\frac{4\pi}{|{\bf x}_3-{\bf x}_2|}\frac{1}{{\beta}^2-4m_e^2\xi^2}\left[e^{-{\beta}|{\bf x}_3-{\bf
x}_2|}-e^{-2m_e\xi|{\bf x}_3-{\bf x}_2|}\right]=
\end{equation}
\begin{displaymath}
=2\pi\Biggl[\frac{\left(1-e^{-2m_e\xi|{\bf x}_3-{\bf x}_2|}\right)}
{2m_e^2\xi^2|{\bf x}_3-{\bf x}_2|}- \frac{{\beta}}{2m_e^2\xi^2}+\frac{\left(1-e^{-2m_e\xi|{\bf x}_3-{\bf
x}_2|}\right){\beta}^2} {8m_e^4\xi^4|{\bf x}_3-{\bf x}_2|}+\frac{{\beta}^2|{\bf x}_3-{\bf x}_2|}{4m_e^2\xi^2}-
\end{displaymath}
\begin{displaymath}
-\frac{{\beta}^3}{8m_e^4\xi^4}-\frac{{\beta}^3({\bf x}_3-{\bf x}_1)^2}{12m_e^2\xi^2}+...\Biggr],
\end{displaymath}
where after an integration the expansion in ${\beta}|{\bf x}_3-{\bf x}_2|$ is used.
For further transformations, it is convenient to use the completeness condition:
\begin{equation}
\label{eq50}
\sum_{n\not =0}\psi_{\mu n}({\bf x}_3)\psi_{\mu n}^\ast({\bf x}_2)=
\delta({\bf x}_3-{\bf x}_2)-\psi_{\mu 0}({\bf x}_3)\psi_{\mu
0}^\ast({\bf x}_2).
\end{equation}

The second and fifth terms in this expansion do not contribute due to the orthogonality of the muon wave functions. 
The first term in square brackets gives the main contribution with respect to $\alpha$ and $W_e/W_\mu$ 
($\gamma=m_e\xi/W_\mu$), which can be split into two parts according to \eqref{eq50}:
\begin{equation}
\label{eq51}
b_{vp,sopt}^{ \mu e}(n\not=0)=b_{vp,1}+b_{vp,2},~~~b_{vp,1}=-\frac{3\alpha^2M_e}{8m_e}\nu_F,
\end{equation}
\begin{equation}
\label{eq52}
b_{vp,2}=\nu_F\frac{\alpha^2M_e}{24\pi m_e}\int_1^\infty\frac{\rho(\xi)d\xi}{\xi}
\frac{[16+\gamma(5\gamma(\gamma+4)+29)]}{(1+\gamma)^4}.
\end{equation}
The total numerical value $b_{vp,1}+b_{vp,2}$ is included in the Table~\ref{tb1}.

Other terms in \eqref{eq48} are calculated as well. With the fourth term in \eqref{eq48}, which is proportional 
to ${\beta}^2=2M_e(E_{\mu n}-E_{\mu 0})$, we can do a number of transformations:
\begin{equation}
\label{eq53}
\sum_{n=0}^\infty E_{\mu n}\int d{\bf x}_2\int d{\bf x}_3\psi_{\mu
0}^\ast({\bf x}_2) \psi_{\mu n}({\bf x}_3)\psi_{\mu n}^\ast({\bf
x}_2)|{\bf x}_3-{\bf x}_2|\psi_{\mu 0}({\bf x}_2)=
\end{equation}
\begin{displaymath}
=\int d{\bf x}_2\int d{\bf x}_3\delta({\bf x}_3-{\bf x}_2)
\left[-\frac{\nabla^2_3}{2M_\mu}|{\bf x}_3-{\bf x}_2|\psi_{\mu 0}^\ast({\bf x}_3)\right]
\psi_{\mu 0}({\bf x}_2).
\end{displaymath}
The expression \eqref{eq53} is divergent due to $\delta$ function. The same divergence take place in another 
term with ${\beta}^2$ in square brackets in \eqref{eq48}. But their sum give finite result:
\begin{equation}
\label{eq54}
b_{{\beta}^2}=\nu_F\frac{9\alpha W_e^2}{32m_eW_\mu}\left(1+\frac{5}{72}\frac{W_\mu^2}{m_e^2}\right).
\end{equation}
Numerically, this correction is significantly smaller than the leading order term in \eqref{eq48}. 
Other terms in \eqref{eq48} can be neglected.

The interaction potential \eqref{eq39} does not contain the muon coordinate. The corresponding contribution 
to the coefficient $c$ in the second order PT can be obtained by setting $n=0$ for the muon state in the 
Coulomb Green's function. Moreover, the presence of $\delta({\bf x}_e)$ in the perturbation operator 
gives the electron Green's function with one zero argument. As a result, the contribution to $c$ 
can be represented in integral form:
\begin{equation}
\label{eq55}
c_{vp,~sopt}^{e N}=\nu_F\frac{\alpha m_\mu g_eg_N}{4\pi m_p}\int_1^\infty\rho(\xi)d\xi
\frac{2\gamma_3^2+3\gamma_3+2\gamma_3\ln \gamma_3-2}{2\gamma_3^3}.
\end{equation}
The vacuum polarization potential in the Coulomb muon-nuclear $(\mu-N)$ interaction does not contribute to $c$ 
in the second order PT due to the orthogonality of the muon wave functions.

Let us consider the calculation of the correction to $c$ from the potential \eqref{eq41} in the second order PT. 
The necessary contribution is determined only by the intermediate muon state with $n=0$ in the Green's function. 
Using \eqref{eq51}, this correction can be represented as:
\begin{equation}
\label{eq56}
c_{vp,~sopt}^{e \mu}=-\nu_F\frac{\alpha m_\mu g_N W_e^2}{6\pi m_p W_\mu^2}\int_1^\infty
\frac{\rho(\xi)d\xi}{(1-\gamma^2)^2}\Bigl[
\frac{3\gamma^2 \gamma_1^2}{(\gamma_1+1)^4}-\frac{\gamma^2}{(\gamma_1+1)^2}-\frac{2\gamma^2 \gamma_1}{(\gamma_1+1)^3}+
\frac{2\gamma_1^2}{(\gamma+\gamma_1)^3}-
\end{equation}
\begin{displaymath}
\frac{2}{(\gamma+\gamma_1)}-\frac{3\gamma_1}{(\gamma+\gamma_1)^2}-\frac{2\gamma_1^2}{(\gamma_1+1)^3}-
\frac{3\gamma_1^2}{(\gamma_1+1)^4}+
\frac{2}{(\gamma_1+1)}+\frac{3\gamma_1}{(\gamma_1+1)^2}+\frac{2\gamma_1}{(\gamma_1+1)^3}-
\end{displaymath}
\begin{displaymath}
\frac{2\gamma_1\ln(\gamma+\gamma_1)}{(\gamma+\gamma_1)^2}+\frac{2\gamma_1(\gamma_1+2-\gamma^2)\ln(\gamma_1+1)}
{(\gamma_1+1)^3}+
\frac{2(\gamma-1)^2\gamma_1(\gamma^2+2\gamma(\gamma_1+1)+\gamma_1^2+3\gamma_1+1)\ln \gamma_1}{(\gamma_1+1)^3
(\gamma+\gamma_1)^2}
\Bigr].
\end{displaymath}

There is also a second-order contribution of PT to the HFS, in which one of the perturbation potentials is determined 
by \eqref{eq32}-\eqref{eq33} (see Fig.~\ref{fig3}), and the second is equal to $\Delta H$. Dividing the correction 
in the hyperfine structure into two parts, we first calculate the part with $n=0$ for the muon ground state. 
The second part with $n\not=0$ contains excited muon states. In turn, the term with $n=0$ can also be divided 
into two parts, and the first part with the $\delta$-function in \eqref{eq32} gives the following contribution 
to $b$:
\begin{equation}
\label{eq57}
b_{vp,~sopt}^{(1)}(n=0)=\nu_F\frac{\alpha}{3\pi}\int_1^\infty\rho(\xi)d\xi
\frac{11W_e}{16 W_\mu}.
\end{equation}

The integral over the spectral parameter $\xi$ is divergent, so we must consider the second term 
in the potential \eqref{eq32}, whose contribution to $b$ will be given by:
\begin{equation}
\label{eq58}
b_{vp,~sopt}^{(2)}(n=0)=\frac{16\alpha^2m_e^2}{9\pi
m_em_\mu}\int_1^\infty \rho(\xi)\xi^2d\xi\int d{\bf x}_3\psi_{e
0}({\bf x}_3)\Delta V_1({\bf x}_3)\times
\end{equation}
\begin{displaymath}
\times\int \sum_{n'\not =0}
\frac{\psi_{e n'}({\bf x}_3)\psi^\ast_{e n'}({\bf x}_1)}{E_{e0}-E_{en'}}\Delta V_2({\bf x}_1)
\psi_{e 0}({\bf x}_1)d{\bf x}_1,
\end{displaymath}
where $\Delta V_1({\bf x}_3)$ is defined by \eqref{eq46} and $\Delta V_2({\bf x}_1)$ \eqref{eq18}. 
Integrating into \eqref{eq58} over all coordinates, we get the following result in leading order in $(W_e/W_\mu)$:
\begin{equation}
\label{eq59}
b_{vp,~sopt}^{(2)}(n=0)=-\nu_F\frac{\alpha m_e}{W_e}\frac{W_e^2}{48\pi
W_\mu^2}\int_1^\infty\rho(\xi)\xi d\xi\frac{32+63\gamma+44\gamma^2+11\gamma^3}{(1+\gamma)^4}.
\end{equation}
This integral also diverges for large values of $\xi$. But the sum of the integrals \eqref{eq57} and \eqref{eq59} 
is finite:
\begin{equation}
\label{eq60}
b_{vp,~sopt}^{(1)}(n=0)+b_{vp,~sopt}^{(2)}(n=0)=\nu_F\frac{\alpha W_e}{48\pi
W_\mu}\int_1^\infty\rho(\xi)d\xi\frac{11+12\gamma+3\gamma^2}{(1+\gamma)^4}.
\end{equation}

Let us proceed to the calculation of terms to $b$ with $n\not=0$. The $\delta$-term in the potential \eqref{eq32}
gives the following contribution:
\begin{equation}
\label{eq61}
b_{vp,~sopt}^{(1)}(n\not =0)=\nu_F\frac{\alpha}{3\pi}\int_1^\infty\rho(\xi)d\xi\left(-\frac{35W_e}{16W_\mu}\right).
\end{equation}
The second term in the potential \eqref{eq32} can be simplified by replacing the exact Green's function 
of the electron with the free one:
\begin{equation}
\label{eq62}
b_{vp,~sopt}^{(2)}(n\not=0)=-\frac{16\alpha^3M_em_e^2}{9\pi m_em_\mu}
\int_1^\infty\rho(\xi)\xi^2d\xi\int d{\bf x}_2\int d{\bf x}_3\times
\end{equation}
\begin{displaymath}
\times\int d{\bf x}_4\psi_{\mu 0}^\ast({\bf x}_4)
\frac{e^{-2m_e\xi|{\bf x}_3-{\bf x}_4|}}{|{\bf x}_3-{\bf x}_4|}\sum_{n\not=0}^\infty
\psi_{\mu n}({\bf x}_4)\psi_{\mu n}({\bf x}_2)|{\bf x}_3-{\bf x}_2|\psi_{\mu 0}({\bf x}_2)
\end{displaymath}
As a result of analytical integration in \eqref{eq62} we get:
\begin{equation}
\label{eq63}
b_{vp,~sopt}^{(2)}(n\not=0)=-\nu_F\frac{\alpha W_e}{3\pi W_\mu}\int_1^\infty\rho(\xi)d\xi[
\frac{1}{\gamma}-\frac{1}{(1+\gamma)^4}(4+\frac{1}{\gamma}+10\gamma+\frac{215\gamma^2}{16}+
\frac{35\gamma^3}{4}+\frac{35\gamma^4}{16})].
\end{equation}
The sum of the contributions \eqref{eq61} and \eqref{eq63} is reduced to
\begin{equation}
\label{eq64}
b_{vp,~sopt}^{(1)}(n\not=0)+b_{vp,~sopt}^{(2)}(n\not=0)=
-\nu_F\frac{\alpha W_e}{3\pi W_\mu}
\int_1^\infty\rho(\xi)d\xi\frac{35+76\gamma+59\gamma^2+16\gamma^3}{16(1+\gamma)^4}.
\end{equation}

Although the absolute values of the calculated vacuum polarization corrections \eqref{eq37}, \eqref{eq43}, 
\eqref{eq44}, \eqref{eq47}, \eqref{eq51}, \eqref{eq60}, \eqref{eq64} are sufficient large, the total contribution 
is small, since the signs of these corrections are different (see Table~\ref{tb1}).

The hyperfine interaction \eqref{eq32} contributes to the coefficient $c$ in the second order PT. Since the muon 
coordinate is not included in \eqref{eq32}, we immediately set $n=0$ for muon intermediate states in 
the Green's function. Then the original formula for this correction is:
\begin{equation}
\label{eq65}
c_{vp,~sopt}=\frac{8\alpha^3 g_N}{9\pi m_e
m_p}\int_1^\infty\rho(\xi)d\xi\int d{\bf x}_1\int d {\bf x}_3\int
d{\bf x}_4|\psi_{\mu 0}({\bf x}_3)|^2\psi^\ast_{e0}({\bf
x}_4)\psi_{e0}({\bf x}_1)\times
\end{equation}
\begin{displaymath}
\times\left[\frac{1}{|{\bf x}_3-{\bf
x}_4|}-\frac{1}{x_4}\right]G_e({\bf x}_4,{\bf
x}_1)\left(\pi\delta({\bf x}_1)-\frac{m_e^2\xi^2}{x_1}e^{-2m_e\xi
x_1}\right).
\end{displaymath}

After analytical integration over ${\bf x}_3$ as in \eqref{eq18} we split \eqref{eq65} into two parts. 
The coordinate integration with the $\delta$-function is done using \eqref{eq30}. In the second term \eqref{eq65} 
we again use the electronic Green's function in the form \eqref{eq23}. The sum of these two terms can be 
expressed in leading order $W_e/W_\mu$ in integral form:
\begin{equation}
\label{eq66}
c_{vp,~sopt}=\nu_F\frac{\alpha g_N m_\mu W_e}{12\pi m_p
W_\mu}\int_1^\infty\rho(\xi)d\xi\frac{3+2\frac{m_e\xi}{W_\mu}}{(1+\frac{m_e\xi}{W_\mu})^2}.
\end{equation}

\section{Nuclear structure and recoil correction}\label{sec5}

\begin{figure}[htbp]
\centering
\includegraphics[scale=0.8]{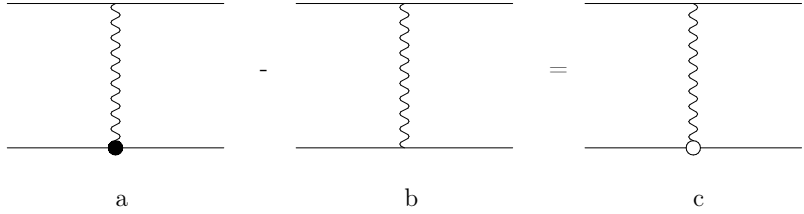}
\caption{Nuclear structure correction to the coefficient $c$ in the $1\gamma$ interaction. 
The bold dot in the diagram represents the nucleus vertex operator. The wavy line denotes 
the hyperfine part of the Breit potential.}
\label{fig:pic3}
\end{figure}

Another class of corrections in the hyperfine structure of muon-electron ions, which is calculated in this work 
to increase the calculation accuracy, is determined by the effects of the structure and recoil of the nucleus 
\cite{apm2004}. We describe the distribution of the charge and magnetic moment of nuclei using the form 
factors $G_E(k^2)$ and $G_M(k^2)$ in the framework of a simple dipole model:
\begin{equation}
\label{eq67}
G_E(k^2)=\frac{1}{\left(1+\frac{k^2}{\Lambda^2}\right)^2},~G_M(k^2)=\frac{G(0)}{\left(1+\frac{k^2}{\Lambda^2}\right)^2},~
G(0)=g_N\frac{m_N}{Zm_p}.
\end{equation}
where the parameter $\Lambda$ is related to the nuclear charge radius $r_N$: $\Lambda=\sqrt{12}/r_N$. In the $1\gamma$ interaction, the correction for the nuclear structure to the coefficient $c$ is determined by the interaction 
amplitude shown in Fig.~\ref{fig:pic3}. The point kernel contribution in Fig.~\ref{fig:pic3}(b) leads to the 
hyperfine splitting \eqref{eq12}. Then the correction for the nuclear structure is determined by the formula:
\begin{equation}
\label{eq68}
c_{str,~1\gamma}=-\nu_F\frac{g_eg_Nm_\mu}{4m_p}\frac{\left(\frac{2W_e}{\Lambda}\right)^3+
3\left(\frac{2W_e}{\Lambda}\right)^2+3\frac{2W_e}{\Lambda}}
{\left(1+\frac{2W_e}{\Lambda}\right)^3}.
\end{equation}

\begin{figure}[htbp]
\centering
\includegraphics[scale=0.8]{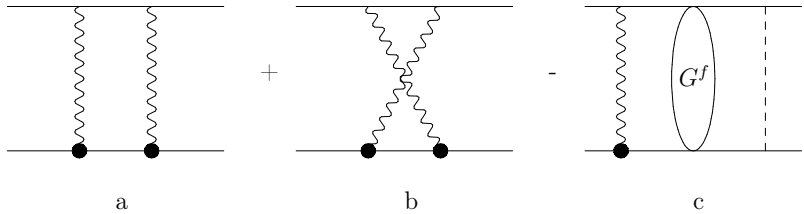}
\caption{Nuclear structure correction to the coefficient $c$ from $2\gamma$ interactions. 
The bold dot in the diagram represents the nucleus vertex operator. The wavy line denotes 
the hyperfine part of the Breit potential. The dotted line corresponds to the Coulomb interaction.}
\label{fig:pic4}
\end{figure}

\begin{figure}[htbp]
\centering
\includegraphics[scale=0.8]{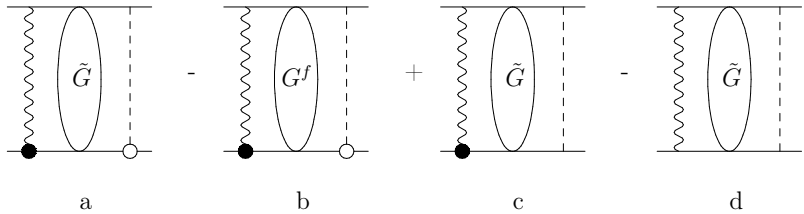}
\caption{Nuclear structure correction to the coefficient $c$ in the second order of PT. 
The bold dot in the diagram represents the nucleus vertex operator. The wavy line denotes 
the hyperfine part of the Breit potential. The dotted line corresponds to the Coulomb interaction. 
$\tilde G$ is the reduced Coulomb Green's function.}
\label{fig:pic5}
\end{figure}

The two-photon amplitudes of the electron-nucleus $(e-N)$ interaction (see Fig.~\ref{fig:pic4}) contribute to a hyperfine splitting of order $\alpha^5$. It can be represented in integral form in terms of the $G_E$ and $G_M$ form factors, 
taking into account the subtractive term \cite{apm2004}:
\begin{equation}
\label{eq69}
c_{str,~2\gamma}=\nu_F\frac{3\alpha M_em_\mu g_e g_N}{2\pi^2 m_p}\int \frac{d{\bf p}}{p^4}\frac{G_M(p)}{G_M(0)}\left[G_E(p)-1\right]=-\nu_F\frac{11Z\alpha M_e m_1g_e g_N}{16 m_p\Lambda},
\end{equation}
where the subtractive term contains the magnetic form factor $G_M(p)$. Integration in \eqref{eq69} is performed 
using the dipole parametrization \eqref{eq67}. Other parts of the iterative term 
$\langle V_{1\gamma}\times G^f\times V_{1\gamma}\rangle_{str}^{hfs}$ are used in the second order 
perturbation theory (see Fig. \ref{fig:pic5}).

\begin{figure}[htbp]
\centering
\includegraphics[scale=0.6]{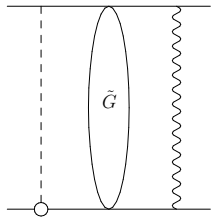}
\caption{Correction to the nucleus structure to $b$ in the second order perturbation theory. 
The wavy line denotes the hyperfine $(e-\mu)$ interaction. $\tilde G$ is the reduced Coulomb Green's function.}
\label{fig:pic6}
\end{figure}

In the second order PT there are two more types of nuclear structure corrections to the coefficient $c$, 
shown in Fig.~\ref{fig:pic5}. The first contribution is determined by the amplitudes shown schematically 
in Fig.~\ref{fig:pic5}(a,b), when the hyperfine part of the perturbation operator is determined by the 
form factor $G_M$, and the second perturbation operator is expressed in terms of the nucleus 
charge radius $r_N$ \cite{apm2015}:
\begin{equation}
\label{eq70}
\Delta V^C_{str, e N}({\bf r})=\frac{2}{3}\pi Z\alpha r^2_N\delta({\bf r}).
\end{equation}
This correction is determined by the following integral expression and can be calculated 
analytically as follows:
\begin{equation}
\label{eq71}
c_{1,str,~sopt}^{e N}=-\nu_F\frac{Z r_N^2 W_e^2g_eg_N m_\mu}{12m_p}\int_0^\infty x^2 dx
e^{-x\left(1+\frac{2W_e}{\Lambda}\right)}\left(-\ln \gamma_1x+\frac{5}{2}-C-\frac{1}{2}\gamma_1x\right)=
\end{equation}
\begin{displaymath}
=-\nu_F\frac{Z r_N^2 W_e^2g_eg_N m_\mu}{12m_p}\frac{[2-\frac{2W_e}{\Lambda}+4(1+\frac{2W_e}{\Lambda})
arccth(1+\frac{4W_e}{\Lambda})]}{(1+\frac{2W_e}{\Lambda})^4}.
\end{displaymath}

Numerically, this contribution $c_{1,str,~sopt}^{e N}$ is proportional to the square of the charge radius of the nuclei 
for which the following values are used: $r(^7_3Li)=2.4440\pm 0.0420 $ fm, $r(^9_4Be )=2.5190\pm 0.0120$ fm, 
$r(^{11}_5B)=2.4060\pm 0.0294$ fm \cite{marinova}.

The correction for the nuclear structure of the second type from the interaction amplitudes in Fig.~\ref{fig:pic5}(c,d) 
is calculated using the potential $\Delta H$ \eqref{eq2} and the nucleus magnetic form factor. In the case 
of the amplitude in Fig.~\ref{fig:pic5}(c), one can perform integration over the muon coordinate in the muon state 
with $n=0$ and over the electron coordinate. After subtracting the point contribution $c_1$, we obtain:
\begin{equation}
\label{eq72}
c_{2,str,~sopt}^{e N}=c_0\frac{4W_e}{(Z-1)\Lambda}\left[
-\frac{3}{2}+\frac{2W_\mu}{\Lambda}-\frac{3}{2}\left(\frac{2W_\mu}{\Lambda}\right)^3+\frac{W_e}{W_\mu}\left(
-\frac{9}{2}+\frac{10W_\mu}{\Lambda}-\frac{25}{6}\left(\frac{2W_\mu}{\Lambda}\right)^2\right)
\right].
\end{equation}

There is nucleus structure contribution to $b$ in the second order perturbation theory, which is shown 
in Fig.~\ref{fig:pic6}. For the Coulomb muon-nucleus interaction, this correction has the form:
\begin{equation}
\label{eq73}
b_{str,sopt}^{\mu N}=\frac{32\pi^2\alpha^2}{3m_em_\mu}r_N^2
\frac{1}{\sqrt{\pi}}\left(W_\mu\right)^{3/2} \int d{\bf
x}_3\psi^\ast_{\mu 0}({\bf x}_3)|\psi_{e0}({\bf x}_3)|^2G_\mu({\bf
x}_3,0,E_{\mu 0})=
\end{equation}
\begin{displaymath}
=-\nu_F \frac{8}{3}W_\mu^2r_N^2\left(\frac{3W_e}{2W_\mu}-
\frac{11}{2}\frac{W_e^2}{W_\mu^2}+\ldots\right),
\end{displaymath}
where the integration result is represented as an expansion in $W_e/W_\mu$.

A structurally similar contribution to $b$ arises from the electron-nuclear interaction. 
It is defined by the following expression:
\begin{equation}
\label{eq74}
b_{str,sopt}^{e N}=\frac{32\pi^2\alpha^2}{3m_em_\mu}r_N^2 \int d{\bf x}_1
\int d{\bf x}_3|\psi^\ast_{\mu 0}({\bf x}_3)|^2\psi_{e0}({\bf
x}_3)G_e({\bf x}_3,{\bf x}_1,E_{e 0}) \psi_{e0}({\bf
x}_1)\delta({\bf x}_1)=
\end{equation}
\begin{displaymath}
=-\nu_F \frac{2W_eW_\mu r_N^2}{(Z-1)}\left[1-\frac{2W_e}{W_\mu}\ln\frac{W_e}{W_\mu}+
\frac{W_e^2}{W_\mu^2}\left(6\ln\frac{W_e}{W_\mu}-4\right)+\ldots\right].
\end{displaymath}
The total nucleus structure correction to $b$, which is equal to the sum of \eqref{eq73} and 
\eqref{eq74}, is included in Table~\ref{tb1}.

\begin{figure}[htbp]
\centering
\includegraphics[scale=0.8]{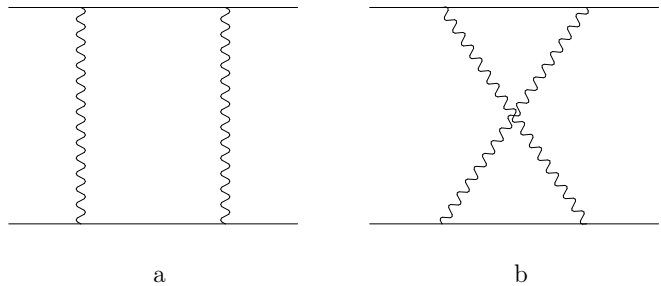}
\caption{Two-photon exchange amplitudes of the electron-muon hyperfine interaction.}
\label{fig:pic7}
\end{figure}

Since the masses of particles in this three-particle system differ greatly from each other, various corrections 
for recoil appear in the calculation of the hyperfine structure, which are determined by the ratio of the masses 
of the particles. Many of the corrections have already been discussed above in the previous sections. 
The interaction operator in a three-particle system is constructed by us as the sum of pair interactions 
that were studied earlier when calculating the fine and hyperfine structure of hydrogen-like atoms 
\cite{egs,be,kp,kks,ns}. The two-photon electron-muon exchange interaction shown in Fig.~\ref{fig:pic7} 
gives a large recoil correction, which is studied in quantum electrodynamics in \cite{egs,ra}. 
The electron-muon interaction operator is defined as follows:
\begin{equation}
\label{eq75}
\Delta V_{rec,\mu e}^{hfs}({\bf x}_{\mu e})=-8\frac{\alpha^2}{m_\mu^2-m_e^2}
\ln\frac{m_\mu}{m_e}({\bf S}_\mu {\bf S}_e) \delta({\bf x}_{\mu e}).
\end{equation}
After averaging the potential \eqref{eq75} over the wave functions \eqref{eq3}, we obtain the contribution 
to the coefficient $b$:
\begin{equation}
\label{eq76}
b_{rec,2\gamma}^{\mu e}=\nu_F\frac{3\alpha}{\pi}\frac{m_em_\mu}{m_\mu^2-m_e^2}
\ln\frac{m_\mu}{m_e}\frac{1}{\left(1+\frac{W_e}{W_\mu}\right)^3}.
\end{equation}

A similar electron-nucleus $2\gamma$ interaction contributes to the coefficient $c$. In the case 
of muonic lithium, beryllium and boron ions, it was studied in \cite{apm2018}. Using the results 
of \cite{apm2018} (see equation (25)), we represent the contribution to $c$ by the following formula:
\begin{equation}
\label{eq77}
c_{rec,2\gamma}^{e N}=-c_0\frac{4Z\alpha m_e}{\pi m_2}\ln\frac{m_2}{m_e}.
\end{equation}
Compared to the main contribution $c_0$, this correction contains two small parameters $\alpha$ and $m_e/m_\mu$, 
but its numerical value slightly increases the accuracy of the result (see Table~\ref{tb1}).

There are also other three-particle two-photon interactions between particles in muon-electron ions. 
So, for example, one photon can give a hyperfine interaction between an electron and a muon, 
and the second - the Coulomb interaction between an electron and a nucleus (or between a muon and a nucleus). 
Assuming that such three-particle amplitudes contribute less to the HFS, we include them in the theoretical 
calculation error.

Let us consider one more correction for nuclear recoil, which is determined by the Hamiltonian $\Delta H_{rec}$ \eqref{eq4}. 
The contribution of $\Delta H_{rec}$ in the second order PT to $c$ is equal to 0, and to the coefficient $b$ 
it is non-zero and is determined by the electron and muon intermediate P-states:
\begin{equation}
\label{eq78}
\Delta b_{rec,sopt}=-\frac{32\pi\alpha^3 M_eM_\mu}{m_em_\mu M_{Li}}\int d{\bf x}_3\int d{\bf x}_2\int d{\bf x}_1
\Psi^\ast_{\mu 0}({\bf x}_3)\Psi^\ast_{e 0}({\bf x}_3)\times
\end{equation}
\begin{displaymath}
\times\sum_{n,n'\not=0}\frac{\Psi_{\mu n}({\bf x}_3)\Psi_{e n'}({\bf x}_3)\Psi^\ast_{\mu n}({\bf x}_2)
\Psi^\ast_{e n'}({\bf x}_1)}{E_{\mu 0}+E_{e 0}-E_{\mu n}-E_{e n'}}({\bf n}_1\cdot {\bf n}_2)
\Psi_{\mu 0}({\bf x}_2)\Psi_{e 0}({\bf x}_1),
\end{displaymath}
where ${\bf n}_1$, ${\bf n}_2$ are unit vectors in coordinate space.

For the analytical calculation \eqref{eq77}, we replace the electron Green's function with the free one, 
as in section 2:
\begin{equation}
\label{eq79}
\Delta b_{rec,sopt}=\frac{16\alpha^3 M^2_eM_\mu}{m_em_\mu M_{Li}}\int d{\bf x}_3\int d{\bf x}_2\int d{\bf x}_1
\Psi^\ast_{\mu 0}({\bf x}_3)\Psi^\ast_{e 0}({\bf x}_3)\times
\end{equation}
\begin{displaymath}
\times\sum_{n\not=0}\Psi_{\mu n}({\bf x}_3)\Psi^\ast_{\mu n}({\bf x}_2)\frac{e^{-b|{\bf x}_3-
{\bf x}_1|}}{|{\bf x}_3-{\bf x}_1|}({\bf n}_1\cdot {\bf n}_2)\Psi_{\mu 0}({\bf x}_2)\Psi_{e 0}({\bf x}_1).
\end{displaymath}
After that, we integrate over ${\bf x}_1$ and expand the result over $b$ (or, which is the same, over $\sqrt{M_e/M_\mu}$):
\begin{equation}
\label{eq80}
\int d{\bf x}_1({\bf n}_1\cdot {\bf n}_2)\frac{e^{-b|{\bf x}_3-{\bf x}_1|}}{|{\bf x}_3-{\bf x}_1|}=
2\pi({\bf n}_2\cdot {\bf n}_3)\left[\frac{4x_3}{3b}-\frac{x^2_3}{2}+\frac{2bx_3^3}{15}+\ldots\right].
\end{equation}
After that, we take the first term in square brackets \eqref{eq79}, perform the integration over the angular variables, and introduce dimensionless variables in the integrals with radial functions:
\begin{equation}
\label{eq81}
\delta b_{rec,sopt}=\nu_F\frac{64M_e}{9M_{Li}}\sqrt{\frac{M_e}{M_\mu}}\sum_{n>1}\frac{n}{\sqrt{n^2-1}}\int_0^\infty x_3^3
R_{10}(x_3)R_{n1}(x_3)dx_3\int_0^\infty x_2^2
R_{10}(x_2)R_{n1}(x_2)dx_2.
\end{equation}
Two contributions from discrete and continuous spectra in \eqref{eq80} have the form:
\begin{equation}
\label{eq82}
\delta b^{(1)d}_{rec,sopt}=\nu_F\frac{2^{11}M_e}{9M_{Li}}\sqrt{\frac{M_e}{M_\mu}}\sum_{n>1}
\frac{n^6(n-1)^{2n-\frac{9}{2}}}{(n+1)^{2n+\frac{9}{2}}}.
\end{equation}
\begin{equation}
\label{eq83}
\delta b^{(1)c}_{rec,sopt}=\nu_F\frac{2^{11}M_e}{9M_{Li}}\sqrt{\frac{M_e}{M_\mu}}\int_0^\infty
\frac{ke^{-\frac{4}{k}arctg(k)}dk}{(1-e^{-2\pi/k})(k^2+1)^{3/2}}.
\end{equation}
The calculation of the second expansion term in \eqref{eq79} gives the following result:
\begin{equation}
\label{eq84}
\delta b^{(2)}_{rec,sopt}=-\nu_F\frac{W_eM_e}{W_\mu M_{Li}},
\end{equation}
which is two orders of magnitude smaller than \eqref{eq81}, \eqref{eq82}.

\section{Electron vertex correction}\label{sec6} 

\begin{figure}[htbp]
\centering
\includegraphics[scale=0.7]{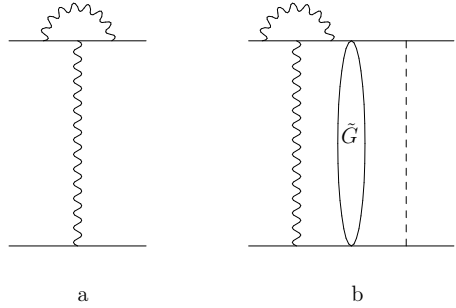}
\caption{Electron vertex correction in the first and second orders of PT. 
The Coulomb photon is represented by a dotted line. The wavy line represents the hyperfine part 
of the Breit potential. $\tilde G$ is the reduced Coulomb Green's function.}
\label{fig:pic8}
\end{figure}

The main contribution of order $\alpha^4$ to the hyperfine structure (the coefficient $b$) is determined 
by the interaction operator \eqref{eq7} as discussed in Section \ref{sec2}. Among different corrections 
to \eqref{eq7} there is a correction determined by the electron vertex function, which is shown 
in Fig.~\ref{fig:pic8}(a). To calculate this contribution, it is first convenient to write it in the 
momentum representation:
\begin{equation}
\label{eq85}
\Delta V^{hfs}_{vert}(k^2)=-\frac{8\alpha^2}{3m_em_\mu}
\left[G_M^{(e)}(k^2)-1\right]
\left({\bf S}_e{\bf S}_\mu\right),
\end{equation}
where $G_M^{(e)}(k^2)$ is the magnetic form factor of the electron, 
and the factor $\alpha/\pi$ is separated from the factor $\left[G_M^{(e)}(k^2)-1\right]$ for convenience.
The commonly used approximation, when the magnetic form 
factor is approximately replaced by its value at zero $G_M^{(e)}(k^2)\approx G_M^{(e)} (0)=1+\kappa_e$, 
is not applicable in this case. Since the typical momentum of an exchange photon is $k\sim\alpha M_\mu$, 
we cannot neglect it in $G_M^{(e)}(k^2)$ as compared to the electron mass $m_e$. Therefore, it is necessary 
to use the exact expression for the Pauli fom factor $g(k^2)$ ($G_M^{(e)}(k^2)-1\approx g(k^2)$) \cite{t4}.

Using the Fourier transform of the potential \eqref{eq85} and averaging it over the wave functions 
\eqref{eq6}, one can represent the electron vertex correction in the HFS as an integral:
\begin{equation}
\label{eq86}
b_{vert,~1\gamma}=\nu_F\frac{\alpha(1+\kappa_\mu)m_e^3W_e}{2\pi^2 W_\mu^4}\int_0^\infty
g(k^2)k^2dk\times
\end{equation}
\begin{displaymath}
\times\left\{\left[1+\left(\frac{m_e}{2W_\mu}\right)^2k^2\right]^2
\left[\left(\frac{W_e}{W_\mu}\right)^2+\left(\frac{m_e}{2W_\mu}\right)^2k^2\right]^2\right\}^{-1},
\end{displaymath}

The contribution \eqref{eq86} is of order $O(\alpha^5 M_e/M_\mu)$. The numerical value \eqref{eq86} is obtained 
after integration over $k$ with a one-loop expression for the form factor $g(k^2)$ \cite{t4} (see the results 
in Table~\ref{tb1}). Using $g(k^2=0)$, we obtain the values of the electron vertex corrections: $41.6139$ MHz $(\mu e Li)$, $140.2879$ MHz $(\mu e Be)$, $332.3111$ MHz $(\mu e B)$ which differ from \eqref{eq86} by approximately 2.5 $\%$.

The contribution of the potential \eqref{eq85} to $b$ in the second order perturbation theory is shown 
in the diagram Fig.~\ref{fig:pic8}(b)). In this case, the second perturbation potential is determined 
by $\Delta H$ \eqref{eq3} (dotted line in the diagram). Let us divide the total contribution of the amplitude 
in Fig.~\ref{fig:pic8}(b) into two parts, which correspond to the muon in the ground state $(n=0)$ and the muon 
in the excited intermediate state $(n\not=0)$ . The first contribution with $n=0$ becomes equal to
\begin{equation}
\label{eq87}
b_{vert,~sopt}(n=0)=\frac{8\alpha^2}{3\pi^2m_em_\mu}\int_0^\infty
k\left[G_M^{(e)}(k^2)-1\right]dk\int d{\bf x}_1\int d{\bf x}_3\psi_{e0}({\bf x}_3)\times
\end{equation}
\begin{displaymath}
\times\Delta\tilde V_1(k,{\bf x}_3)G_e({\bf x}_1,{\bf x}_3)
V_\mu({\bf x}_1)\psi_{e0}({\bf x}_1),
\end{displaymath}
where $V_\mu({\bf x}_1)$ is determined by \eqref{eq18}, and
\begin{equation}
\label{eq88}
\Delta \tilde V_1(k,{\bf x}_3)=\int d{\bf x}_4\psi^\ast_{\mu 0}({\bf x}_4)
\frac{\sin(k|{\bf x}_3-{\bf x}_4|)}{|{\bf x}_3-{\bf x}_4|}
\psi_{\mu 0}({\bf x}_4)=\frac{\sin(kx_3)}{x_3}\frac{1}{\left[1+\frac{k^2}{(2W_\mu)^2}\right]^2}.
\end{equation}

After substituting the electron Green's function \eqref{eq27} into \eqref{eq87}, we reduce this expression 
to an integral form:
\begin{equation}
\label{eq89}
b_{vert,~sopt}(n=0)=\nu_F\frac{\alpha}{2\pi^2}\left(\frac{m_e}
{W_\mu}\right)^2\left(\frac{W_e}{W_\mu}\right)^2\int_0^\infty
\frac{k\left[G_M^{(e)}(k^2)-1\right]dk}{\left[1+\frac{m_e^2k^2}{(2W_\mu)^2}\right]^2}\times
\end{equation}
\begin{displaymath}
\times\int_0^\infty x_3e^{-\frac{W_e}{W_\mu}x_3}\sin\left(\frac{m_e k}{2W_\mu}x_3\right)dx_3
\int_0^\infty x_1\left(1+\frac{x_1}{2}\right)e^{-x_1\left(1+\frac{W_e}{W_\mu}\right)}
dx_1 \Biggl[\frac{W_\mu}{W_e x_>}-\ln(\frac{W_e}{W_\mu}x_<)-
\end{displaymath}
\begin{displaymath}
-\ln(\frac{W_e}{W_\mu}x_>)+Ei(\frac{W_e}{W_\mu}x_<)+\frac{7}{2}-2C-\frac{W_e}{W_\mu}\frac{(x_1+x_3)}
{2}+\frac{1-e^{\frac{W_e}{W_\mu}x_<}}{\frac{W_e}{W_\mu}x_<}\Biggr].
\end{displaymath}
All integrations over the coordinates $x_1$, $x_3$ are performed analytically, and over $k$ numerically. 
The intermediate expression before integration over $k$ is omitted, since it has a cumbersome form.

The second part of the vertex correction (Fig.~\ref{fig:pic8}(b)) with $n\not=0$ after a series of simplifications 
can be transformed into
\begin{equation}
\label{eq90}
b_{vert,~sopt}(n\not=0)=\nu_F\frac{W_e W_\mu^3}{\pi^3(Z-1)}\int e^{-W_\mu x_2}
d{\bf x}_2\int e^{-W_e x_3}d{\bf x}_3\int e^{-W_\mu x_4}d{\bf x}_4\times
\end{equation}
\begin{displaymath}
\times\int_0^\infty k\sin(k|{\bf x}_3-{\bf x}_4|)\left(G_M^{(e)}(k^2)-1\right)
\frac{|{\bf x}_3-{\bf x}_2|}{|{\bf x}_3-{\bf x}_4|}\left[\delta({\bf x}_4-{\bf x}_2)-
\psi_{\mu 0}({\bf x}_4)\psi_{\mu 0}({\bf x}_2)\right].
\end{displaymath}

The contributions of the two terms in square brackets \eqref{eq90} will be presented separately after integration 
over the coordinates ${\bf x}_1$ and ${\bf x}_3$ ($\gamma_2=m_ek/2W_\mu$):
\begin{equation}
\label{eq91}
b^{(1)}_{vert,~sopt}(n\not=0)=\nu_F\frac{\alpha}{2\pi^2}\left(\frac{m_e}{
W_\mu}\right)^3\frac{W_e}{(Z-1)W_\mu}\int_0^\infty
k^2\left[G_M^{(e)}(k^2)-1\right]dk\frac{1}{(\gamma_1^2-1)^3}\times
\end{equation}
\begin{displaymath}
\times\left[\frac{4 \gamma_1(\gamma_1^2-1)}{(1+\gamma_2^2)^3}-\frac{\gamma_1
(3+\gamma_1^2)}{(1+\gamma_2^2)^2}+\frac{4\gamma_1^2(\gamma_1^2-1)}{(\gamma_1^2+
\gamma_2^2)^3}+\frac{1+3\gamma_1^2}{(\gamma_1^2+\gamma_2^2)^2}\right],
\end{displaymath}
\begin{equation}
\label{eq92}
b^{(2)}_{vert,~sopt}(n\not=0)=-\nu_F\frac{\alpha}{2\pi^2}\left(\frac{m_e}{
W_\mu}\right)^3\frac{W_e}{(Z-1)W_\mu}\int_0^\infty k^2\left[G_M^{(e)}(k^2)-1\right]dk\times
\end{equation}
\begin{displaymath}
\times\frac{1}{(1+\gamma_2^2)^2}\left[\frac{2}{(\gamma_1^2+\gamma_2^2)}-\frac{(\gamma_1+1)}
{[(1+\gamma_1)^2+\gamma_2^2]^2}-\frac{2}{(\gamma_1+1)^2+
\gamma_2^2}-\frac{\gamma_2^2-3\gamma_1^2}{(\gamma_1^2+\gamma_2^2)^3}\right].
\end{displaymath}

Note that the theoretical error in the sum of contributions $b^{(1)}_{vert,~sopt}(n\not =0)+b^{(2)}_{vert,~sopt}(n\not =0)$ 
is determined by the factor $\sqrt {M_e/M_\mu}$ connected with the omitted terms in the expansion of the form 
\eqref{eq24}. 
It can be about $10\%$ of the total \eqref{eq91}-\eqref{eq92} result, which is represented 
by a separate line in Table~\ref{tb1}.

The considered electron vertex corrections in hyperfine splitting are of order $\alpha^5$. The total value 
of the resulting vertex contribution (see Table~\ref{tb1}) differs from the above values in the approximation 
when the form factor $g(k^2)$ is replaced by the anomalous magnetic moment of the electron.

\begin{table}[htbp]
\caption{\label{t1} Contributions to the coefficients $b$, $c$ in the hyperfine splitting of the ground state 
in lithium, beryllium and boron ions, and helium atoms. Table rows in order correspond to 
$(\mu e _3^7 Li)^+$, $(\mu e _4^9 Be)^{2+}$, $(\mu e _5^11 B)^{3+}$, $(\mu e _2^3 He)$, $(\mu e _2^4 He)$}
\bigskip
\label{tb1}
\begin{ruledtabular}
\begin{tabular}{|| c | c | c | c ||}  \hline
Contribution                     &  $b$, MHz &    $c$, MHz     & Reference           \\ 
to the coefficients $b$ and $c$  &           &                 &                     \\ \hline
Leading order contribution                & 35830.53  & 4422.90         & \eqref{eq11},\eqref{eq12}\\ 
of order $\alpha^4$              & 120791.04 &-5397.57         &                          \\ 
                                 & 286127.05 &29216.41         &                          \\ 
                                 & 4487.7131 &-1083.3208       &                          \\   
                                 & 4488.6167 &   0             &                     \\  \hline
Recoil correction    &-155.58     &22.27             &\eqref{eq28}, \eqref{eq31}, \eqref{a7} \\
of order $\alpha^4\frac{W_e}{W_\mu}$, $\alpha^4\frac{W^2_e}{W^2_\mu}$ &-390.95   &-20.36  &     \\
                                      & -738.06    &88.11   &                          \\ 
                                      & -29.9789   &-8.3125  &                          \\ 
                                      & -29.7371   &0  &                          \\ \hline                                                                            
One-loop VP correction                &0.70        &0.17    &\eqref{eq37},\eqref{eq38} \\
in $1 \gamma $ interaction            &3.99        &-0.32   &                          \\
                                      &13.59       &2.29    &                          \\
                                      &0.0357      &0.0272   &                          \\ 
                                      &0.0359      &0       &                          \\ \hline                          
One-loop VP correction                &0.69        &    0     &\eqref{eq44}              \\
in $\mu N$ interaction                &3.15        &    0     &                          \\
in second order of PT                 &9.01        &    0     &                          \\
                                      &0.0485      &   0 &                          \\ 
                                      &0.0484      &   0 &                          \\ \hline  
One-loop VP correction                &-0.86       &-0.05   &\eqref{eq47},\eqref{eq51},\\
in $\mu e$ interaction                &-3.07       &0.07    &\eqref{eq54},\eqref{eq56} \\
in second order of PT                 &-7.73       &-0.39   &                          \\
                                      &-0.1010     &0.0090   &                          \\ 
                                      &-0.1012     &0   &                          \\ \hline    
\end{tabular}
\end{ruledtabular}
\end{table}

\begin{table}[htbp]
Table \ref{tb1} (continued)
\bigskip
\begin{ruledtabular}
\begin{tabular}{|| c | c | c | c ||}  \hline
One-loop VP correction                &1.14        &0.27    &\eqref{eq43},\eqref{eq55} \\
in $ e N$ interaction                 &5.85        &-0.45   &                          \\ 
in second order of PT                 &19.00       &3.09    &                          \\ 
                                      &0.0752      &-0.0440   &                        \\ 
                                      &0.0756      & 0  &                          \\ \hline 
One-loop VP correction                &-0.50       & 0.04   &\eqref{eq60}, \eqref{eq64}, \eqref{eq66}\\
with $\Delta H$ potential             &-1.52       &-0.05   &             \\ 
in second order of PT                 &-3.28       & 0.23   &                          \\ 
                                      &-0.0704     &-0.0129   &                          \\ 
                                      &-0.0732     & 0  &                          \\ \hline                                       
Nuclear structure correction          &   0        & -0.71  &\eqref{eq68}              \\
in $1 \gamma$ interaction             &   0        &  1.33  &                          \\ 
                                      &   0        & -9.19  &                          \\ 
                                      &   0        & 0.0696   &                          \\ 
                                      &   0        &  0 &                          \\ \hline                                                                           
Nuclear structure correction          &     0      &    -0.49      &\eqref{eq69}              \\
in $2 \gamma$ interaction             &   0        &   0.82       &                          \\ 
                                      &   0        &    -5.27      &                          \\ 
                                      & 0          &  0.0638     &                          \\ 
                                      & 0          & 0       &                          \\ \hline                                                                               
Nuclear structure correction          & -0.47      & -0.35  &\eqref{eq71}, \eqref{eq72}, \eqref{eq73}, \eqref{eq74}  \\
in second order of PT                 & -3.30      & 0.44   &                  \\ 
                                      & -11.77     & -2.30  &                          \\
                                      & -0.0132    & 0.0690   &                          \\ 
                                      & -0.0097    & 0    &                          \\ \hline 
Nuclear recoil correction             & 0.53       & 0 &\eqref{eq82}, \eqref{eq83}, \eqref{eq84}\\
from $\Delta H_{rec}$                 & 1.38       & 0   &        \\ 
                                      & 2.68       & 0  &                          \\ 
                                      & 0.1078     & 0  &                          \\ 
                                      & 0.0809     & 0  &                          \\ \hline  
\end{tabular}
\end{ruledtabular}
\end{table}
\begin{table}[htbp]
Table \ref{tb1} (continued)
\bigskip
\begin{ruledtabular}
\begin{tabular}{|| c | c | c | c ||}  \hline
Electron vertex correction            & 40.96     & 0  &\eqref{eq86}\\
of order $\alpha^5$                     & 136.73    & 0  &  \\ 
in $1\gamma$ interaction                & 320.59    & 0 &                          \\
                                        & 5.1765    & 0  &                          \\
                                        & 5.1774    & 0  &                          \\ \hline                                        
Electron vertex correction            & -0.06     & 0  &\eqref{eq91},\eqref{eq92},\eqref{eq89}\\
of order $\alpha^5$                     & -0.06     & 0  &  \\ 
in second order of PT                   & 0.02      & 0 &                          \\ 
                                        & -0.0209    & 0  &                          \\ 
                                        & -0.0206    & 0  &                          \\ \hline                                        
Recoil correction                       & 6.43       & -0.09  &\eqref{eq76},\eqref{eq77}\\
in $2\gamma$ interaction                & 21.68      & 0.12   &  \\ 
                                        & 51.35      & -0.67  &                          \\ 
                                        &  0.8055    & 0.0315    &                          \\
                                        & 0.8056     & 0  &                          \\ \hline                               
Relativistic correction                 & 5.77       & 1.41  &\eqref{eq93}\\
of order $\alpha^6$                     & 53.05      & -3.88   &  \\ 
                                        & 241.24     & 37.30  &                          \\ 
                                        & 0.0401     & -0.0864  &                          \\ 
                                        & 0.0401     & 0  &                          \\ \hline                                                                            
Radiative correction                    & -3.48       & -0.64  &\eqref{eq94}\\
of order $\alpha^6$                     & -11.74      & 1.04   &  \\ 
                                        & -27.82      & -7.02  &                          \\ 
                                        & -0.4345     & 0.1041   &                          \\ 
                                        & -0.4346     & 0  &                          \\ \hline                                                                            
Summary values                        & 35725.80   & 4444.73  &                          \\ 
                                      & 120606.23  & -5418.81 &                          \\ 
                                      & 285995.87  & 29322.59 &                          \\ 
                                      & 4463.3835  & -1091.4024&                          \\ 
                                      & 4464.5042  &   0      &                          \\ \hline                                                                            
\end{tabular}
\end{ruledtabular}
\end{table}

\section{Conclusion}\label{Conc} 

In this paper, we calculate the intervals of the hyperfine structure of the ground state for muon-electron 
ions of lithium, beryllium, boron, and helium using the perturbation theory method formulated earlier for muonic 
helium ions in \cite{lm1,lm2}. To increase the accuracy of calculations, we took into account corrections 
in the hyperfine structure of orders $\alpha^5$ and $\alpha^6$, connected with the effects of vacuum polarization, 
the nucleus structure and recoil, and electron vertex corrections. All obtained numerical results are presented 
in the Table~\ref{tb1}. It specifies the correction values for lithium, beryllium and boron ions with an accuracy 
of two decimal places, and for muonic helium with an accuracy of four decimal places. This is due to the increase 
in the total value of contributions due to the nuclear charge Z during the transition from muon-electron helium 
to boron.

Let us note the main features of the performed calculations:
\begin{enumerate}
\item Muon-electron ions of lithium, beryllium and boron have a complex hyperfine structure in the ground state, 
which arises as a result of the interaction of the magnetic moments of the nucleus, electron and muon. We have 
explored small intervals of the hyperfine structure that can be measured.
\item When calculating the HFS, there are small parameters of the fine structure constant and the particle mass 
ratio, which can be used in constructing expansions in perturbation theory. In this paper, corrections of order $\alpha^4$, 
$\alpha^5$, and $\alpha^6$ are considered, taking into account the recoil effects of the first and second orders.
\item Vacuum polarization effects are of great importance for achieving high accuracy in the calculation of hyperfine 
splitting. They lead to a modification of the two-particle interaction potentials, which give corrections of order 
$\alpha^5\frac{M_e}{M_\mu}$. We take into account the contribution of one-loop vacuum polarization in the first 
and second orders of perturbation theory.
\item The electron vertex correction to the coefficient $b$ is obtained taking into account the one-loop expression 
for the magnetic form factor of the electron, since the characteristic momentum entering the vertex operator is 
of order of the electron mass.
\item Corrections for the structure of the nucleus are expressed both in terms of electromagnetic form factors and 
in terms of the charge radius.
\item Relativistic corrections to the coefficients $b$ and $c$ are obtained using expressions from \cite{huang}:
\begin{equation}
\label{eq93}
b_{rel}=\left(1+\frac{3}{2}(Z-1)^2\alpha)^2-\frac{1}{3}(Z\alpha)^2\right)\nu_F,~~~
c_{rel}=\frac{3}{2}(Z-1)^2\alpha^2 c_0.
\end{equation}
\item To estimate the radiative corrections without recoil of the order of $O(\alpha^6)$ in the HFS, we use the 
results of analytical calculations in two-particle atoms \cite{be,kp,kks,ns}, which give the following expressions 
for $b$ and $c$:
\begin{equation}
\label{eq94}
b_{\alpha^2}=\alpha^2\nu_F(\ln 2-\frac{5}{2}),~~~
c_{\alpha^2}=\frac{1}{2}\alpha(Z\alpha)(\ln 2-\frac{5}{2})c_0.
\end{equation}
\end{enumerate}

Using the total numerical values for the coefficients $b$ and $c$ presented in the Table~\ref{tb1}, we obtain 
the following values for the hyperfine intervals for lithium, beryllium and boron\eqref{eq17}: 
$\Delta\nu_1(\mu e^7_3Li)$=13994.76 MHz,
$\Delta\nu_1(\mu e^9_4Be)$=85539.16 MHz, 
$\Delta\nu_1(\mu e^{11}_5B)$=123767.56 MHz,
and 
$\Delta\nu_2(\mu e^7_3Li)$=21731.04 MHz,
$\Delta\nu_2(\mu e^9_4Be)$=35067.07 MHz, 
$\Delta\nu_2(\mu e^{11}_5B)$=162228.31 MHz,

In the case of muonic helium, the hyperfine splitting of the ground state has the form:
$\Delta\nu(\mu e^3_2He)=\frac{3}{4}(b-c)=4166.089$ MHz,
$\Delta\nu(\mu e^4_2He)=4464.504$ MHz.

These numerical values agrees with the experimental data \eqref{eq2}, taking into account the available 
theoretical and experimental errors. Our results are also in good agreement with recent calculations 
using the variational method in \cite{korobov}: 4166.39(58) MHz $(\mu e^3_2 He)$, 4464.55(60) MHz $(\mu e ^4_2 He)$.

Previously, the calculation of hyperfine intervals in the muonic lithium-7 ion was performed in \cite{af1,af2} 
within the framework of the variational method. Our results generally agree with \cite{af2} results on 
muon-electron lithium-7: $\Delta\nu_1(\mu e^7_3Li)$=13989.19 MHz, $\Delta\nu_2(\mu e^7_3Li)$=21729.22 MHz. 
A slight difference is due to the inclusion in our work of corrections for the vacuum polarization and the structure 
of the nucleus. In the case of muon-electron helium, the results \cite{af2} of the hyperfine splitting 
of the ground state $\Delta\nu(\mu e^3_2He)$=4166.383 MHz, $\Delta\nu(\mu e^4_2He)$=4464.554 MHz differ 
from our values by approximately 0.29 MHz $(\mu e^3_2He)$ and 0.05 MHz $(\mu e^4_2He)$.

We performed an analytical calculation of recoil corrections of orders
$\frac{W_e^2}{W_\mu^2}\ln \frac{W_e}{W_\mu}$, $\frac{W_e^2} {W_\mu^2}$.
As already noted, in \cite{lm1} the recoil corrections \eqref{eq24} were calculated numerically for muonic helium-4. 
The sum of \eqref{eq20}-\eqref{eq21} contributions obtained in \cite{lm1} is (-29.65) MHz. In our work, a similar 
contribution is determined by the sum of (-29.8306) MHz and 0.0935 MHz (\eqref{a7}) 
(see Table~\ref{tb1}) and is 
equal to (-29.7371) MHz, which differs from the result \cite{lm1} by 0.087 MHz. 

An analysis of individual contributions to the hyperfine structure coefficients $b$ and $c$ in Table~\ref{tb1} shows 
that relativistic corrections, corrections for nucleus structure and recoil, vacuum polarization, and electron vertex 
corrections must be taken into account to achieve good calculation accuracy. The theoretical uncertainty can be estimated 
in terms of the Fermi energy $\nu_F$ and the parameters $W_e$ and $W_\mu$. The main source of the theoretical error 
is recoil corrections of order $(W_e/W_\mu)^{5/2}\ln(W_e/W_\mu) \nu_F$, which are not always taken into account 
exactly in the calculations. So for muonic helium-3,4 the error is about 0.008 MHz, for lithium - 0.13 MHz, 
for beryllium - 0.56 MHz, for boron - 1.53 MHz.

\begin{acknowledgments}
The work was supported by the Foundation for the Advancement of Theoretical Physics and Mathematics “BASIS" 
(Grant № 19-1-5-67-1 (F.A. Martynenko))
\end{acknowledgments}

\appendix

\section{The estimation of other recoil contributions to the coefficient (24)}\label{app} 

As noted in section \ref{sec2}, the contribution to \eqref{eq21} is calculated using the approximation 
of the free Green's function for the electron $G_e^0$. The following term $G_e^0V^CG_e^0$ in the expansion 
of the Green's function contributes to the coefficient $b_1(n\not=0)$ in \eqref{eq21} of the form:
\begin{equation}
\label{a1}
b_1^{(2)}(n\not=0)=-\frac{\alpha W_e^2 g_e g_\mu}{3\pi (Z-1)m_e m_1}|\psi_{e0}(0)|^2
\int \psi_{\mu 0}({\bf x}_3)d{\bf x}_3 \int \psi_{\mu 0}({\bf x}_2)d{\bf x}_2
\sum_{n\not=0}\psi_{\mu n}({\bf x}_3) \psi_{\mu n}({\bf x}_2)\times
\end{equation}
\begin{displaymath}
\int \frac{d{\bf x}}{|{\bf x}|}\int \frac{d{\bf x}_1}{|{\bf x}_2-{\bf x}_1|}
\frac{e^{-\beta|{\bf x}-{\bf x}_3|}}{|{\bf x}-{\bf x}_3|}
\frac{e^{-\beta|{\bf x}-{\bf x}_1|}}{|{\bf x}-{\bf x}_1|}.
\end{displaymath}

After calculating the integral over ${\bf x}_1$, we use the expansion over $\beta|{\bf x}-{\bf x}_2|$, 
as in \eqref{eq24}. The first term $1/\beta$ of the expansion does not contribute due to the orthogonality 
of the muon wave functions, while the second term gives the following correction:
\begin{equation}
\label{a2}
b_1^{(2)}(n\not=0)=\frac{2\alpha W_e^2 g_e g_\mu}{3 (Z-1)m_e m_1}|\psi_{e0}(0)|^2
\int \psi_{\mu 0}({\bf x}_3)d{\bf x}_3 \int \psi_{\mu 0}({\bf x}_2)d{\bf x}_2\times
\end{equation}
\begin{displaymath}
\sum_{n\not=0}\psi_{\mu n}({\bf x}_3) \psi_{\mu n}({\bf x}_2)I({\bf x}_2,{\bf x}_3),~~~
I({\bf x}_2,{\bf x}_3)=\int d{\bf x} \frac{|{\bf x}-{\bf x}_2|}{|{\bf x}|}
\frac{e^{-\beta|{\bf x}-{\bf x}_3|}}{|{\bf x}-{\bf x}_3|}.
\end{displaymath}

Let us expand the integral $I({\bf x}_2,{\bf x}_3)$ into a series for small values of $x_2^i$:
\begin{equation}
\label{a3}
I({\bf x}_2,{\bf x}_3)=I(0)+x_2^iI_i(0)+\frac{1}{2}x_2^i x_2^j I_{ij}(0),~~~
I_i(0)=\frac{dI}{dx_2^i}|_{x_2^i}=0,~~~
I_{ij}(0)=\frac{d^2I}{dx_2^i dx_2^j}|_{x_2^i}=0.
\end{equation}

Using the exact form of $I({\bf x}_2,{\bf x}_3)$, let us calculate $I_i(0)$, $I_{ij}(0)$ and obtain:
\begin{equation}
\label{a4}
I({\bf x}_2,{\bf x}_3)=\frac{8\pi}{\beta^2}-({\bf x}_2{\bf x}_3)\frac{4\pi}{9}
\bigl[-4+3C+\frac{3}{2}\ln\frac{M_e}{M_\mu}+3\ln (W_\mu x_3)+\frac{3}{2}\ln\frac{n^2-1}{n^2}\bigr]+
\end{equation}
\begin{displaymath}
\frac{\pi}{6}\Bigl[x_2^2\ln(W_\mu x_3)-\frac{({\bf x}_2{\bf x}_3)^2}{x_3^2}\Bigl(
11-9C-9\ln(W_\mu x_3)-\frac{9}{2}\ln\frac{M_e}{M_\mu}\Bigr)\Bigr].
\end{displaymath}

Separating the terms dependent and independent of $n$ and calculating the corresponding matrix elements 
in the same way as in section \ref{sec2}, we obtain the following additional correction to the coefficient 
$b_1(n\not=0)$:
\begin{equation}
\label{a5}
b_1^{(2)}(n\not=0)=-\nu_F\frac{W_e^2}{3(Z-1)W_\mu^2}\frac{g_e}{g_\mu}{4}
\left[\frac{7}{4}+\frac{3}{2}\ln\frac{M_e}{4M_\mu}+S_{ln}^d+S_{ln}^c\right],
\end{equation}
\begin{equation}
\label{a6}
S_{ln}^d=2^{11}\sum_{n>1}\ln\frac{n^2-1}{n^2}\frac{n^7(n-1)^{2n-5}}{(n+1)^{2n+5}},~~~
S_{ln}^c=2^{11}\int_0^\infty\ln(k^2+1)\frac{kdk}{(k^2+1)^5(1-e^{-\frac{2\pi}{k}})}e^{-\frac{4}{k}\arctan k}.
\end{equation}

The numerical values of the correction \eqref{a5} for the considered ions are:
\begin{equation}
\label{a7}
b_1^{(2)}(n\not=0)=
\begin{cases}
0.65~MHz,~~~\mu e ^7_3 Li \\
1.84~MHz,~~~\mu e ^9_4 Be \\
3.71~MHz,~~~\mu e ^11_5 B \\
0.0951~MHz,~~~\mu e ^3_2He \\
0.0935~MHz,~~~\mu e ^4_2He \\
\end{cases}.
\end{equation}

They are taken into account when obtaining the total result in the Table~\ref{tb1}.

\end{document}